\documentclass[prd,showpacs,altaffilletter,twocolumn,
superscriptaddress,floatfix,nofootinbib]{revtex4} 

\usepackage{amsfonts,amsmath,graphicx,bm,siunitx}
\usepackage{dcolumn}
\usepackage[pdftex,
       colorlinks=true,
       pdftitle={The Hindered M$1$ Radiative Decay
         $\Upsilon(2S)\to\eta_b(1S)\gamma$ from Lattice NRQCD},
       pdfpagemode=None,
       bookmarksopen=true]{hyperref}

\usepackage{ifpdf}
\usepackage{multirow}
\usepackage[colorlinks=true]{hyperref}
\usepackage{url}

\newcommand{\cambridge}{Department of Applied Mathematics and
Theoretical Physics, University of Cambridge, Cambridge CB3 0WA, UK}

\newcommand{\glasgow}{SUPA, School of Physics and Astronomy,
  University of Glasgow, Glasgow, G12 8QQ, UK}

\newcommand{\mainz}{Institut f{\"u}r Kernphysik, University of Mainz,
  Becherweg 45, 55099 Mainz, Germany}

\newcommand{\VEC}[1]{{\bf \bm{#1}}} 

\def\today{\number\day\space\ifcase\month\or
January\or February\or March\or April\or May\or June\or
July\or August\or September\or October\or November\or December\fi
\space\number\year}
\newcount\mins \newcount\hours
\def\now{\hours=\time \mins=\time
	\divide\hours by60 \multiply\hours by60 \advance\mins by-\hours
	\divide\hours by60 
	\number\hours:\ifnum\mins<10 0\fi\number\mins }

\begin{document}

\title{Hindered M1 Radiative Decay of 
         $\bm{\Upsilon(2S)}$ from Lattice NRQCD }

\author{Ciaran \surname{Hughes}} 
\email[]{ch558@cam.ac.uk}
\affiliation{\cambridge}

\author{Rachel~J.~\surname{Dowdall}} 
\affiliation{\cambridge}

\author{Christine~T.~H.~\surname{Davies}} 
\affiliation{\glasgow}

\author{Ronald~R.~\surname{Horgan}} 
\affiliation{\cambridge}

\author{Georg von \surname{Hippel}} 
\affiliation{\mainz}

\author{Matthew \surname{Wingate}}
\affiliation{\cambridge}

\collaboration{HPQCD Collaboration}
\email[]{http://www.physics.gla.ac.uk/HPQCD}


\begin{abstract}
We present a calculation of the hindered M$1$  $\Upsilon(2S)\to\eta_b(1S)\gamma$   decay rate using lattice non-relativistic QCD. The calculation includes spin-dependent relativistic corrections to the NRQCD action through $\mathcal{O}(v^6)$ in the quark's relative velocity, relativistic corrections to the leading order current which mediates the transition through the quark's magnetic moment, radiative corrections to the leading spin-magnetic coupling and for the first time a full error budget. We also use gluon field ensembles at multiple lattice spacing values, all of which include $u$, $d$, $s$ and $c$ quark vacuum polarisation. Our result for the branching fraction is $\mathcal{B}(\Upsilon(2S)\to\eta_b(1S)\gamma) = 5.4(1.8)\times 10^{-4} $, which agrees with the current experimental value.
\end{abstract}

\maketitle



\section{Introduction}

Quantum Chromodynamics (QCD) has been accepted as the theory describing the strong force of nature ever since the discovery of the $J/\psi$. Since then, there has been
a long history of using the spectrum and decays of heavy quarkonia in
order to understand QCD, heavy quarkonia being the ideal theoretical
testing grounds when using potential models, and more recently, lattice QCD. Heavy quarkonium states below threshold are very narrow, and electromagnetic transition rates are therefore significant. Comparing the theoretical and experimental rates for these decays then provides a very clear test of our understanding of the internal structure of heavy quarkonia. 

A certain class of electromagnetic transitions between quarkonium
states, known as hindered M$1$ transitions, require a spin-flip
between different radial excitations and are particularly sensitive
to small relativistic effects \cite{Godfrey:2001} which can illuminate the
dynamics of the initial and final state systems. These
hindered M$1$ transitions still remain a challenge from both
the experimental and theoretical perspective. Within the bottomonium sector, such decays
include the $\Upsilon(2S)\to\eta_b(1S)\gamma$ radiative transition, where BaBar measured
$\mathcal{B}(\Upsilon(2S)\to\eta_b(1S)\gamma) = 3.9(1.5)\times
10^{-4}$ \cite{BaBar:Upsilon2S} in $2009$.

On the theory side, hindered M$1$ decays have been nortoriously
difficult to pin down from within a potential model framework 
\cite{Godfrey:2001}, where systematic errors are hard to quantify and branching fractions ranging from $0.05\times 10^{-4}$ to $15 \times 10^{-4}$ are found. The reasons for the difficulty in accurately predicting these
decays from within a potential model will be discussed in Section \ref{sec:Conclusions}. The continuum effective field theory approach called potential NRQCD (pNRQCD) has been used to predict radiative bottomonium decays, including M$1$ transitions. While these calculations have become quite precise for the allowed $1$S $\to$ $1$S M$1$ transitions, the results for hindered M$1$ transitions are dominated by theoretical uncertainties and presently can only give an order-of-magnitude estimate \cite{Brambilla:M1, Segovia:M1}.

Lattice NRQCD is a first principles tool that has been systematically
improved by the HPQCD collaboration and can aid in reliably pinning
down this difficult to predict decay. Using this formalism, one can
accurately  overcome each of the issues arising from within
a potential model framework. Previous exploratory work on this decay in a
lattice NRQCD framework was done in \cite{Lewis:Rad1, Lewis:Rad2}. 
We make a  number of improvements to those studies so that an accurate
calculation can be done, complete with a full error budget. Some
of these improvements include using one-loop radiative corrections
in the NRQCD action and we show in Section \ref{sec:DecayNRQCD} that these decays are
very sensitive to a subset of these radiative corrections. 

This paper is organised as follows. In Section \ref{sec:DecayRate} we
set up notation and formulae relevant to this decay, and in Section
\ref{sec:CompDetails} we give details of the computational
setup including a discussion of states in NRQCD at non-zero momentum. In Section \ref{sec:Currents} the different currents mediating
this transition in NRQCD are shown and the perturbative calculation of the matching
coefficient from the leading order current to full QCD is performed. Finally, analysis of the $\Upsilon(2S)\to\eta_b(1S)\gamma$
decay rate with a full error budget is given in Section \ref{sec:DecayNRQCD}. We conclude with a discussion in Section \ref{sec:Conclusions}.


\section{Decay Rates for Radiative Transitions}
\label{sec:DecayRate}
BaBar has measured the branching
fraction of the $\Upsilon(2S)\to\eta_b(1S)\gamma$ decay as $3.9(1.5) \times
10^{-4}$ \cite{BaBar:Upsilon2S}, which when combined with the $\Upsilon(2S)$ total width
$31.98\pm 2.63$ keV \cite{PDG:2014}, gives the decay rate $1.25(49) \times
10^{-2}$ keV.  The large errors on the branching fraction are due to
the difficulty in isolating the small $\eta_b(1S)$ signal from other
nearby photon lines ($\chi_{bJ}(2P,1P)\to\Upsilon(1S)\gamma$,
$\Upsilon(3S,2S)\to \Upsilon(1S)\gamma$) and from the large background
in the energy spectrum of inclusive decays \cite{QWG:2010}.

We want to perform an accurate and reliable theoretical
calculation to compare to this experimental result. Computation of the theoretical decay
rate requires the matrix element of the appropriate operator between the $\Upsilon(2S)$ and $\eta_b(1S)$ states as
input. In a Lorentz invariant theory, using the
fact that the matrix element transforms as a vector under parity (and
parity invariance of our theory), the
only possible decomposition of the matrix element is 
\begin{align}
& \langle \eta_{b(mS)}(k) | j^{\mu}(0) | \Upsilon_{({nS})}(p,
 \epsilon(p,\lambda)) \rangle  =\nonumber \\ 
& \hspace{1cm}  \frac{2\mathcal{V}^{\Upsilon\eta_b}_{nm}(q^2)}{m_{\Upsilon(nS)} + m_{\eta_b(mS)}}
\varepsilon^{\mu\nu\rho\sigma}p_{\nu}k_{\rho}\epsilon(p,\lambda)_{\sigma} \label{eqn:MatrixElement}
\end{align}
where $q$ is the photon momentum, $\epsilon(p,\lambda)_{\sigma}$ is the polarisation vector of the $\Upsilon_{(nS)}$ and $p = k  + q$  by momentum
conservation.  Using time reversal invariance, one can show that
$\mathcal{V}^{\Upsilon\eta_b}_{nm}(q^2)$ is real \cite{Dudek:CharmRad}. As the $\Upsilon(2S)$ is a $\bar{b}b$ bound state, this M$1$ (spin-flip) transition can occur by flipping the spin on either the quark or the antiquark. Since this is a symmetric process, the form factor resulting from coupling the current to the quark or to the anti-quark is then identical. In our lattice calculation we
only couple the current to the quark (c.f.
Sec.~\ref{sec:Currents}) and actually compute
$V^{\Upsilon\eta_b}_{nm}(q^2)|_{\text{lat}} = \mathcal{V}^{\Upsilon\eta_b}_{nm}(q^2) /2$ . 

The decay rate can now be written as
\begin{align}
& \Gamma(\Upsilon(2S)\to\eta_b(1S)\gamma) = \nonumber \\
& \hspace{0.5cm}  \frac{16\alpha_{\text{QED}}
  e^2_q}{3}\frac{|\VEC{q}|^3}{(m_{\Upsilon(2S)} +
  m_{\eta_b(1S)})^2}|V^{\Upsilon\eta_b}_{21}(0)|_{\text{lat}}|^2 \label{eqn:UpsilonDecayRate} 
\end{align}
where $\alpha_{\text{QED}}$ is the fine structure constant, $e_q$  is the quark charge in units of $e$ (i.e., $-1/3$ for $b$-quarks) and $|\VEC{q}| = (m^2_{\Upsilon(2S)} - m^2_{\eta_b(1S)})/
2m_{\Upsilon(2S)}$ by energy conservation, ensuring that the photon is on-shell with $q^2= 0$. Thus, from the theoretical
perspective, the most challenging part of calculating the decay rate from
first principles is computing the single
unknown dimensionless hadronic form factor $\mathcal{V}^{\Upsilon\eta_b}_{21}(q^2=0)$, which encodes the
nonperturbative effects of QCD. This quantity can be
calculated in lattice QCD, and this study will focus on
the computation of $V^{\Upsilon\eta_b}_{21}(q^2=0)|_{\text{lat}}$. 

Using the experimental value of the decay rate mentioned above, as well as $|\VEC{q}| = 609(5)$ MeV measured
from experiment \cite{BaBar:Upsilon2S} and $\alpha_{\text{QED}}= 1/137$, we
infer
\begin{align}
V^{\Upsilon\eta_b}_{21}(q^2 =0)|_{\text{exp}} & = 0.069 (14) \,. \label{eqn:FormFactorExp}
\end{align}
This form factor can be directly compared to $V^{\Upsilon\eta_b}_{21}(q^2=0)|_{\text{lat}}$. From now on, we will drop the $|_{\text{lat}}$ subscript to avoid superfluous notation.


\section{Computational Details}
\label{sec:CompDetails}

\subsection{Second Generation $N_f = 2+1+1$ Gluon Ensembles}
\label{sec:Ensembles}

Our calculation uses gauge field configurations generated by the MILC collaboration \cite{MILC:Configs}. For the gauge fields, they used the tadpole-improved L\"{u}scher-Weisz gauge action, fully improved to $\mathcal{O}(\alpha_sa^2)$. This is possible
as the gluon action has coefficients corrected perturbatively through
$\mathcal{O}(\alpha_s)$, including pieces proportional to the number
of quark flavours in the sea \cite{hart:GluonImprovement}. These ensembles are said to
have $2+1+1$ flavours in the sea, the up and down quarks (treated as two degenerate light quarks with mass $m_l$), the strange quark, and the charm quark. The sea  quarks are included using the HISQ formulation of fermions \cite{HISQAction}, fully improved to 
$\mathcal{O}(\alpha_sa^2)$, removing one-loop taste-changing processes and possessing smaller discretisation errors compared to the previous staggered actions.

Five ensembles were chosen, spanning three lattice spacing and three values of
$m_l/m_s$, so that any dependence on the lattice spacing and sea quark mass
could be fit and extrapolated to the physical limit. Details are given in Table
\ref{tab:GluonEnsembles}. Due to the computational expense, most of the ensembles use heavier $m_l$ than in the real world; however one of the ensembles used in this study (set $4$ in Table \ref{tab:GluonEnsembles}) has physical $am_l/am_s$, enabling our calculations to be performed at the physical  point and reducing uncertainties associated with unphysically heavy sea quark masses. 

Successive configurations generated within each ensemble are expected to be
correlated. These autocorrelations in meson correlators were studied in
\cite{Dowdall:Upsilon} for
the ensembles in Table \ref{tab:GluonEnsembles}. There we find that the
autocorrelations for bottomonium correlators are not appreciable and that
the configurations can be treated as statistically independent. The ensembles have been fixed to Coulomb gauge to allow
non-gauge invariant smearings to be used, helping extract precise
results for the excited states in our calculation (c.f.\ Sec.~\ref{sec:2pt}).

\begin{table}[t]
  \caption{Details of the gauge ensembles used in this
    study. $\beta$ is the gauge coupling. $a_{\Upsilon}$ is the lattice spacing determined from the
    $\Upsilon(2S-1S)$ splitting \cite{Dowdall:Upsilon}, where the
    error combines statistics, experiment and the dominant NRQCD
    systematic error. $am_q$ are the sea
    quark masses, $N_s \times N_T$ gives the spatial and temporal
    extent of the lattices in lattice units and $n_{\text{cfg}}$  is the number of
    configurations in each ensemble. We use $16$ time sources on each configuration to increase statistics. Ensemble $1$ is     referred to as ``very coarse'', 2, 3, and 4 as ``coarse,'' and 5 as 
    ``fine''. }
\label{tab:GluonEnsembles}
\begin{center}
  \begin{tabular}{l l l l l l l l }
    \hline \hline 
    Set&  $\beta$ & $a_{\Upsilon}$(fm) &  $am_l$ &$am_s$ & $am_c$ & $N_s
    \times N_T$ & $n_{\text{cfg}}$ \\ \hline
    $1$ & $5.8$ & $0.1474(15)$ & $0.013$ & $0.065$ & $0.838$ &$16 \times
    48$ & $1020$ \\ \hline
    $2$ & $6.0$ &$0.1219(9)$ & $0.0102$ & $0.0509$ & $0.635$ &$24 \times
    64$ & $1052$ \\
    $3$ & $6.0$ &$0.1195(10)$ & $0.00507$ & $0.0507$ & $0.628$ &$32 \times
    64$ & $1000$ \\
    $4$ & $6.0$ & $0.1189(9)$ & $0.00184$ & $0.0507$ & $0.628$ &$48 \times
    64$ & $1000$ \\ \hline
    $5$ & $6.3$ &$0.0884(6)$ & $0.0074$ & $0.037$ & $0.440$ &$32 \times
    96$ & $1008$ \\ \hline \hline
  \end{tabular}
\end{center}
\end{table}

\subsection{$b$-quarks Using NRQCD}
\label{sec:NRQCD}

This study focuses purely on bottomonium processes, and information
on these processes can be
computed on the lattice using combinations of $b$-quark propagators,
calculated on
the gluon ensembles listed in Table \ref{tab:GluonEnsembles}. As the $b$-quark
has a Compton wavelength of about $0.04$ fm, these lattices
cannot resolve relativistic $b$-quark formulations, owing to $a
> 0.08$ fm. However, it is well known that $b$-quarks are very
nonrelativistic inside their bound states ($v^2\approx 0.1$), and thus, using a nonrelativistic effective field theory (NRQCD) for bottomonium
states is very appropriate. Within NRQCD, with expansion parameter
$v$ (the velocity of the quark inside the bound state), one writes down a tower of
operators to a certain order in $v$ allowing for a systematic inclusion of ever-decreasing relativistic corrections. This effective field theory is then discretised as lattice
NRQCD \cite{Lepage:ImprovedNRQCD}. There are a number of systematic
improvements which need to be 
made in order to produce highly accurate results. These will be addressed shortly. 

We use a lattice NRQCD action correct through $\mathcal{O}(v^4)$, with
additional spin-dependent $\mathcal{O}(v^6)$ terms\footnote{The
  quantities relevant to this study are insensitive to the spin-independent
  $\mathcal{O}(v^6)$ terms within our precision.} and include 
discretisation corrections.
This lattice formalism has already been used successfully to study
bottomonium $S$, $P$ and $D$ wave mass splittings \cite{Dowdall:Upsilon,
  Daldrop:Dwave}, precise hyperfine 
splittings \cite{Dowdall:Hyperfine, Dowdall:Hl}, $B$ meson decay
constants \cite{Dowdall:BMeson}, $\Upsilon$ and $\Upsilon'$ leptonic
widths \cite{Brian:LeptonicWidth} and $B$, $D$ meson mass
splittings \cite{Dowdall:Hl}. The Hamiltonian evolution equations
can be written as
\begin{align}
G({\VEC{x}},t+1)  & = e^{-aH}  G({\VEC{x}},t) \nonumber \\
G({\VEC{x}},t_{\text{src}})  & = \phi({\VEC{x}})
\end{align}
with 
\begin{align}
e^{-aH} & = \left( 1 - \frac{a\delta H|_{t+1}}{2} \right) \left( 1
  - \frac{aH_0|_{t+1}}{2n} \right)^n U_t^{\dagger}(x) \nonumber \\
&\hspace{0.7cm}  \times \left( 1 -  \frac{aH_0|_t}{2n} \right)^n \left( 1 - \frac{a\delta H|_t}{2}
\right) \label{eqn:NRQCDGreensFunction}  \\
aH_0  & = -\frac{\Delta^{(2)}}{2am_b}, \nonumber \\
 a\delta H  & =  a\delta H_{v^4} + a\delta H_{v^6}; \nonumber  
\end{align}
\begin{align}
a\delta H_{v^4} & = -c_1 \frac{(\Delta^{(2)})^2}{8(am_b)^3} 
+c_2\frac{i}{8(am_b)^2}\left( {\VEC{\nabla \cdot \tilde{E}}} -
  {\VEC{\tilde{E} \cdot \nabla  }} \right)  \nonumber \\ &
-c_3\frac{1}{8(am_b)^2} {\VEC{\sigma \cdot \left( \tilde{\nabla}
      \times \tilde{E} - \tilde{E} \times \tilde{\nabla} \right) }}
\nonumber \\ &
-c_4\frac{1}{2am_b} {\VEC{ \sigma \cdot \tilde{B}}} 
+c_5 \frac{\Delta^{(4)}}{24am_b}
-c_6 \frac{(\Delta^{(2)})^2}{16n(am_b)^2} \nonumber 
\end{align}
\begin{align}
a\delta H_{v^6} & = 
-c_7\frac{1}{8(am_b)^3} {\VEC{ \left\{ \Delta^{(2)} , \sigma \cdot
      \tilde{B} \right\} }}  \nonumber \\ & 
-c_8\frac{3}{64(am_b)^4} {\VEC{ \left\{ \Delta^{(2)}, \sigma \cdot \left( \tilde{\nabla}
      \times \tilde{E} - \tilde{E} \times \tilde{\nabla}
    \right)\right\} }}  \nonumber \\ &
-c_9 \frac{i}{8(am_b)^3} {\VEC{\sigma \cdot \tilde{E} \times
    \tilde{E} }} ~. \label{eqn:LNRQCDv6} 
\end{align}
The parameter $n$ is used to prevent instabilities at large
momentum due to the kinetic energy operator. A value of $n=4$ is
chosen for all $am_b$ values. A smearing function 
$\phi({\VEC{x}})$ is used to improve projection
onto a particular state in the lattice data. Using an array of smearing functions to
improve the overlap with the ground state and the first excited state
will prove crucial to obtaining accurate results for the
$\Upsilon(2S)\to\eta_b(1S)\gamma$ decay. To evaluate the propagator, we use random wall sources that are
implemented stochastically with $U(1)$ 
white noise, significantly improving the precision of the S-wave
states \cite{Dowdall:Upsilon}.

Here, $am_b$ is the bare $b$ quark mass, $\nabla$ is the symmetric
lattice derivative, with $\tilde{\nabla}$ the improved version, and
$\Delta^{(2)}$, $\Delta^{(4)}$ are the lattice discretisations of 
$\Sigma_i D_i ^2$, $\Sigma_i D_i^4$ respectively. ${\VEC{\tilde{E}}}$,
${\VEC{\tilde{B}}}$ are the improved chromoelectric and chromomagnetic
fields, details of which can be found in \cite{Dowdall:Upsilon}. Each of these fields, as
well as the covariant derivatives, must be tadpole-improved using the same improvement procedure as in the perturbative calculation of the matching coefficients \cite{Hammant:2013,Dowdall:Upsilon}
(thus removing unphysical tadpole diagrams from using the Lie group
element rather than the Lie algebra element in the construction of the
lattice field theory). We take the mean trace of the gluon field in
Landau gauge, $u_{0L} = \langle \frac{1}{3}\text{Tr}\,U_{\mu}(x)\rangle$, as the
tadpole parameter, calculated in \cite{Dowdall:Upsilon, Dowdall:BMeson}. 

The matching coefficients $c_i$ in the above Hamiltonian take into
account the high-energy UV modes from QCD processes that are not
present in NRQCD.  Each $c_i$ can be expanded perturbatively as
$c_i = 1 + c_i^{(1)}\alpha_s + \mathcal{O}(\alpha_s^2) $ and, after tadpole
improvement, we expect the radiative corrections $c_i^{(1)}$ to be
$\mathcal{O}(1)$. Each $c_i^{(1)}$ can then be fixed by matching a particular lattice
NRQCD formalism\footnote{Changing the NRQCD action can modify the
  Feynman rules used in the computation of $c_i^{(1)}$ in perturbation
  theory, in general changing its value.} to full continuum QCD. These corrections have
previously been computed \cite{Hammant:2013, Dowdall:Upsilon}. Alternatively, particular
$c_i$'s can be tuned nonperturbatively, which we discuss in Section \ref{sec:HyperTest}. 

A high-precision calculation with a reliable error budget will require
knowledge of at least the $\mathcal{O}(\alpha_s)$ corrections to the
matching coefficients. For example, when tuning the quark mass
$am_b$ fully nonperturbatively in NRQCD, one computes the kinetic mass
of a hadron\footnote{The static mass (the energy corresponding to zero-spatial momentum) in lattice NRQCD  \cite{Dowdall:Upsilon} is shifted due to
  the removal of the mass term from the Hamiltonian and so one can only
  tune static mass differences fully
  nonperturbatively.} \cite{Dowdall:Upsilon}. This kinetic mass depends on the internal kinematics
of the hadron, and hence on the terms $c_1$, $c_5$, and $c_6$ in the
Hamiltonian. Using the one-loop corrected coefficients to these terms
has a small but visible effect on the kinetic masses and
hence on the value of the tuned $am_b$ \cite{Dowdall:Upsilon}. 

In addition to this, for an $\mathcal{O}(v^4)$ NRQCD action with $c_4=1$, the kinetic mass for
the $\eta_b$ is actually found to be larger than that of the $\Upsilon$ \cite{Dowdall:Upsilon}, opposite to
what is seen at zero momentum and, more importantly, in experiment. The
explanation is that the $\sigma \cdot B$ term gives rise to the
hyperfine splitting, and the splitting from this term is correctly included in
the static mass (the mass at zero energy, offset due to removing the
mass term from the Lagrangian). However, relativistic corrections
to $\sigma \cdot B$ (the
term proportional to $c_7$ in the Hamiltonian above) are needed to
correctly feed this splitting into the kinetic mass. On a fine
lattice, a value of $c_4=1.18$ and $c_7=1.25$ 
was needed to yield a hyperfine splitting using kinetic masses which
agreed with experiment within 
errors \cite{Dowdall:Hyperfine}. In order to remove the sensitivity to the
$\sigma \cdot B$ term when tuning $am_b$, one does not use the kinetic
mass of a single state, but the spin-averaged kinetic mass of the
$\Upsilon$  and $\eta_b$ \cite{Dowdall:Upsilon, Meinel}. Including $a\delta H_{v^6}$ terms
in the evolution equations makes the $\eta_b$ kinetic mass lower than that of the $\Upsilon$,
as they include relativistic corrections to
the $\sigma \cdot B$ term. The spin-averaged kinetic mass gets smaller and
the bare quark mass gets larger \cite{Dowdall:Hyperfine}.

 The parameters used in this study are
summarised in Table \ref{tab:NRQCDParams}. There, $c_1,c_5$ and $c_6$ are the correct values for a $v^4$ NRQCD action \cite{Dowdall:Upsilon}, but the small changes to these coefficients in going to a $v^6$ NRQCD action have a negligible effect on the quantities studied here, as shown in Figure \ref{fig:Varyci}. While the $am_b$ values
from ensembles $1,2$ and $5$  listed in Table 
\ref{tab:NRQCDParams} have all been tuned against the spin-averaged
kinetic mass using the Hamiltonian above \cite{Dowdall:Hyperfine}, the
$am_b$ values from ensembles
$3$ and $4$ were previously tuned without the $a\delta H_{v^6}$ terms \cite{Dowdall:BMeson}.  Ensembles $2,3$ and
$4$ are all coarse lattices and only differ by having different light quark masses in
the sea. Ensemble $2$ has a correctly tuned $am_b=2.73$ for the Hamiltonian
we use, corresponding to $m_b=4.418$ GeV. It is appropriate to tune the
$am_b$ values on the other coarse lattices to match this physical
value. Using the lattice spacings listed in Table
\ref{tab:GluonEnsembles}, we find the $am_b$ values on ensemble $3$
and $4$ listed in Table \ref{tab:NRQCDParams}.  All these ensembles
have essentially the same value of the 
lattice spacing, so the running of the bare mass is a
negligible effect. This was observed with a $\mathcal{O}(v^4)$ Hamiltonian \cite{Dowdall:Upsilon}.

\begin{table}[t]
  \caption{Parameters used for the valence quarks. $am_b$ is the bare
    $b$-quark mass in lattice units, $u_{0L}$ is the tadpole
    parameter. The $c_i$ are coefficients of terms in the NRQCD
    Hamiltonian (see Eq. \ref{eqn:LNRQCDv6}). Details of their calculation can be found in \cite{Hammant:2013, Dowdall:Upsilon}. $c_3,c_7,c_8$ and $c_9$ are included at tree-level. We also list the values of $\alpha_s$ used to determine the one-loop corrections in the perturbative matching in Sec.~\ref{sec:MatchingCoeff} and for the error budget in Sec.~\ref{sec:fullerror}.
   \label{tab:NRQCDParams}}
\begin{center}
  \begin{tabular}{l l l l l l l l  }
    \hline \hline 
    Set&  $am_b$ & $u_{0L}$ &  $c_1$, $c_6$ & $c_2$ &  $c_4$ & $c_5$ & $\alpha_s(\pi / a)$ \\ \hline
    $1$ & $3.31$ & $0.8195$ & $1.36$ & $1.29$ &  $1.23$ & $1.21$ &
    $0.275$ \\ \hline
    $2$ & $2.73$ & $0.8346$ & $1.31$ & $1.02$ & $1.19$ & $1.16$ &
     $0.255$ \\ 
    $3$ & $2.68$ & $0.8349$ & $1.31$ & $1.02$ & $1.19$ & $1.16$ &
    $0.255$ \\ 
    $4$ & $2.66$ & $0.8350$ & $1.31$ & $1.02$ & $1.19$ & $1.16$ &
     $0.255$ \\ \hline
    $5$ & $1.95$ & $0.8525$ & $1.21$ & $0.68$ & $1.18$ & $1.12$ &
    $0.225$ \\ 
 \hline \hline
  \end{tabular}
\end{center}
\end{table}

Within NRQCD, the Dirac field $\Psi$ can be written in terms of the quark
$\psi$ and anti-quark $\chi$ as $\Psi = (\psi, \chi)^T$. The propagator
is then found to be
\begin{center}
$S(x|y) =
 \begin{pmatrix}
  G_{\psi}(x|y) & 0 \\
  0 & -G_{\chi}(x|y) 
 \end{pmatrix}
$
\label{eqn:DiracMatixNRQCD}
\end{center}
where $G_{\psi}(x|y)$ is the two-spinor component quark propagator and
$G_{\chi}(x|y)$ is the two-spinor component anti-quark
propagator. $\gamma^5$ hermicity becomes $G_{\psi}(x|y) = -
G_{\chi}^{\dagger}(y|x)$. As such, we write our interpolating
operators as in Table \ref{tab:Operators} and
then use the above decomposition, with suitable boundary conditions, to write the correlator
in terms of $G_{\psi}(x|y)$. 

\begin{table}[t]
  \caption{The local bilinear operators used in this study. Note the
    $i\gamma^5$ is needed to make the overlaps real \cite{Dudek:CharmRad}. The second
    column gives the $J^{PC}$ states that these operators create at
    rest in an infinite volume continuum. The
    third column gives the helicity eigenvalues $\lambda$ that these
    operators create at nonzero momentum in an infinite volume
    continuum which is only rotationally invariant, while the $J$ in brackets
    are the states which contribute to that helicity (c.f.
    Section \ref{sec:EnergyEigenstates}). }
\label{tab:Operators}
\begin{center}
  \begin{tabular}{l l l}
    \hline\hline
    $\mathcal{O}^{\Gamma}(x)$ & $J^{PC}$ & \multicolumn{1}{l}{$\lambda (\leftarrow J^P)$}
    \\ \hline

    $\bar{\psi} i\gamma^5 \psi$ & $0^{-+}$ & $0^- (\leftarrow
    J^P = 0^-,1^+,2^-,\ldots)$ \\ \hline

    \multirow{2}{*}{$\bar{\psi} \gamma^i \psi$} &
    \multirow{2}{*}{$1^{--}$ } & $0^+ (\leftarrow J^P =
    0^+,1^-,2^+,\ldots)$ \\ 
    & & $|1| (\leftarrow J = 1,2,3,\ldots )$ \\\hline\hline
  \end{tabular}
\end{center}
\end{table}

\subsection{Non-Integer Momentum on the Lattice}
\label{sec:Mom}
Using periodic boundary conditions (PBC) for the quark fields forces the momentum components to be $p_i = 2\pi n_i
/L$, where $n_i$ is an integer. The issue with this is that processes which
occur at a specific momentum, such as that needed for an on-shell photon
in the form factor $V^{\Upsilon\eta_b}_{21}(q^2=0)$, cannot be reached at an
integer-valued momentum. Here, we use ``twisted boundary conditions''
($\theta$BC) \cite{TwistedBC, TwistedBC1} in order to find the matrix element at the physical $q^2=0$
point. There are some subtleties with using $\theta$BC in our calculation
that, to our knowledge, are not found in the literature, and we give an explicit
example of the construction of our twisted correlators in Appendix
\ref{app:TwistedCorrelators}. As seen there, and confirmed by numerical data, the twisted and untwisted correlator data should agree (if the same momentum is used) on a configuration level if everything is
done correctly. 

In our calculations, we choose ${\VEC{p_i}} = {\VEC{p_f}} = {\VEC{q}} ={\VEC{0}}$ and
only twist a single propagator so that ${\VEC{p_f^{\theta}}} =
    -{\VEC{q^{\theta} =\theta}}$. The choice of isotropic twist momentum ${\VEC{\theta}} = \chi_0(1,1,1)\times 2\pi /L $ that gives $q^2=0$ depends on the
specific process under study and for the $\Upsilon(2S)\to \eta_b(1S)\gamma$
decay $\chi_0$ is found from (\ref{eqn:UpsilonDecayRate}) as:
\begin{align}
 \chi_0 & = \frac{L}{2\sqrt{3}\pi} \frac{m_{\Upsilon(2S)}^2 -
  m_{\eta_b(1S)}^2}{2m_{\Upsilon(2S)}} \label{eqn:TwistValues}
\end{align}
yielding $|{\VEC{q}}^{\theta}|^2 = |{\VEC{\theta}}|^2$. We
choose an isotropic momentum  as it has been
shown to 
reduce discretisation errors from rotational symmetry breaking \cite{Dowdall:Upsilon}. 
Since static masses obtained from correlators at rest are shifted by an
arbitrary value in NRQCD, tuning $\chi_0$ from lattice
data would require a more lengthy computation of the kinetic masses. Instead, we use the experimental values of these masses \cite{PDG:2014} to tune $\chi_0$ and check that $q^2=0$ from the results.

\subsection{Energies and Amplitudes from Lattice QCD}
\label{sec:2pt}

Extracting matrix elements on the lattice requires knowledge of the
lattice amplitudes and energies corresponding to the states being studied. The lattice quantity which most naturally
encodes information on these is the two-point correlator
\begin{align}
& C_{\text{2pt}}(n_{src}, n_{snk}, {\VEC{p^{\theta}}} ;t) = \nonumber \\ 
& \hspace{1cm} \sum_{{\VEC{x}}} e^{-i {\VEC{x\cdot p^{\theta}}}} \langle \mathcal{O}(n_{snk}; {\VEC{x}},t+t_0)  \mathcal{O}^{\dagger}(n_{src}; {\VEC{0}},t_0)  \rangle \label{eqn:2ptCorr}
\end{align}
Here, $t_0$ is the source time, $n_{src}, n_{snk}$ are the smearing
type (discussed below) and ${\VEC{p^{\theta}}}$ is the
twisted momentum. After performing
the Wick contractions with the bilinear operators listed in Table \ref{tab:Operators},
the connected\footnote{Disconnected diagrams for heavy quarkonia are expected to be negligible as they are suppressed by the heavy quark mass \cite{Dudek:CharmRad}.} correlator has the form
\begin{align}
& C_{\text{2pt}}(n_{src}, n_{snk}, {\VEC{p^{\theta}}};t) = \nonumber \\ 
& \hspace{1cm} \sum_{{\VEC{x}}} e^{-i {\VEC{x\cdot p^{\theta}}}} \text{Tr} \left[  \Gamma_{src} S(0|x; n_{src};n_{snk}) \Gamma_{snk} \tilde{S}^{\theta}(x|0) \right] \nonumber
\end{align}
where $\tilde{S}^{\theta}$ is the twisted propagator (c.f. Appendix
\ref{app:TwistedCorrelators}). We use smearing functions $\phi^{src}(r),
\phi^{snk}(r)$ on the anti-quark field at the source and sink respectively.
We employ hydrogen-like wavefunctions which have been successful in previous studies of $b$-physics: $\phi(r) =$ $\delta_{r,0}$, $\exp(-r/r_0)$,
$(2r_0-r)\exp(-r/2r_0)$. $r_0$ is the smearing radius, and we point the
reader to \cite{Dowdall:Upsilon} for further details on the
smearings\footnote{We use the smearing types $l,g,e$ as described in
  that reference.}. The different smearing combinations
used in this study give a $3\times 3$ matrix of correlators.  We do not smear the quark fields due to 
complications on using twisted-smeared fields as outlined in Appendix
\ref{app:TwistedCorrelators}.

The two-point correlator in (\ref{eqn:2ptCorr}) can be spectrally decomposed as
\begin{align}
C_{\text{2pt}}(n_{src},n_{snk}, {\VEC{p^{\theta}}} ;t)  = \sum_{{{k=1}}}^{n_{\text{exp}}} a(n_{snk},k)a^*(n_{src},k) e^{-E_kt} \label{eqn:2ptCorrFit}
\end{align}
where $E_k$ is the $(k-1)^{\text{th}}$ energy excitation of the interpolating operator
$\mathcal{O}(x)$ used in the construction of the correlator
and $a(n_{src}/n_{snk},k)$ are the corresponding amplitudes,
labelled by the smearing used at the source or sink. We are only
interested in the first few excited states, so we do not need to
worry about multiparticle states or the open $b$-threshold. Our two-point correlators are propagated for a maximum of $t/a=15$ timeslices, as after this the locally smeared correlator on a fine lattice is largely saturated by the ground state. In addition, correlators were calculated with $16$ different
 time sources on each configuration in order to increase
statistics. To avoid complications due to correlations between these time sources, 
correlators were then averaged over all sources on the same configuration. 

We fit  the $3\times 3$  matrix of
correlators from $t/a =1-15$  using a simultaneous multi-exponential
Bayesian fit \cite{Lepage:Code, Lepage:Fitting} to the
spectral decomposition in (\ref{eqn:2ptCorrFit}). Different smearings give rise to different amplitudes and so we take priors on them
to be $0.1(1.5)$. The priors on
the ground state energies are estimated from previous results and given a
suitably wide width \cite{Dowdall:Upsilon}. For the zero momentum
case, prior information tells us that the energy splittings $E_{n+1} -
E_n$ are of the order $500(250)$ MeV, while for the nonzero momentum case, priors of $480(250)$ MeV are used (due to the inclusion of additional states in the correlator, see Sec.~\ref{sec:EnergyEigenstates}). Logarithms of the energy splittings  are taken in the fit to ensure that the
ordering of states is preserved, helping the stability of the fit
\cite{Lepage:Fitting}.

\subsection{Energy Eigenstates in Lattice NRQCD}
\label{sec:EnergyEigenstates}

\begin{figure}[t]
  \centering
  \includegraphics[width=0.49\textwidth]{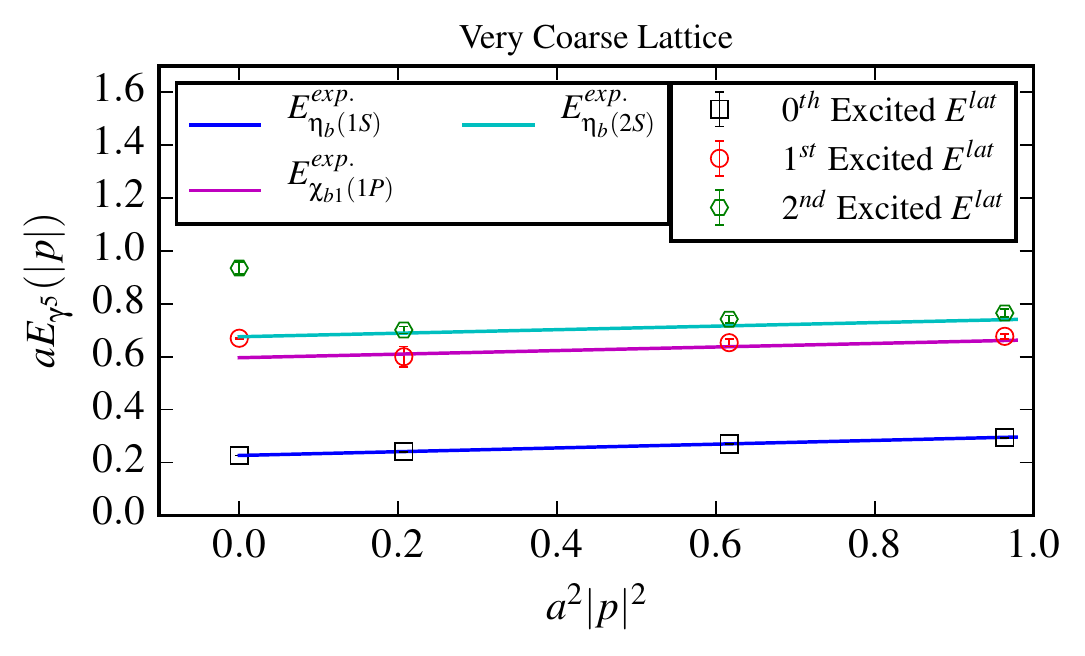}
  \includegraphics[width=0.49\textwidth]{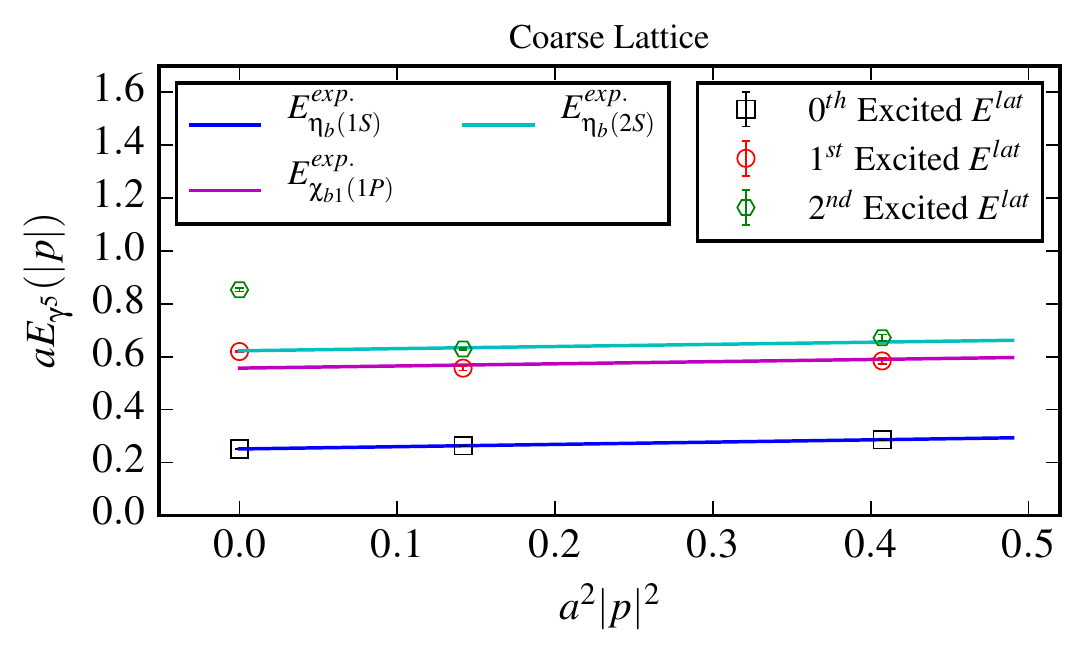}
  \includegraphics[width=0.49\textwidth]{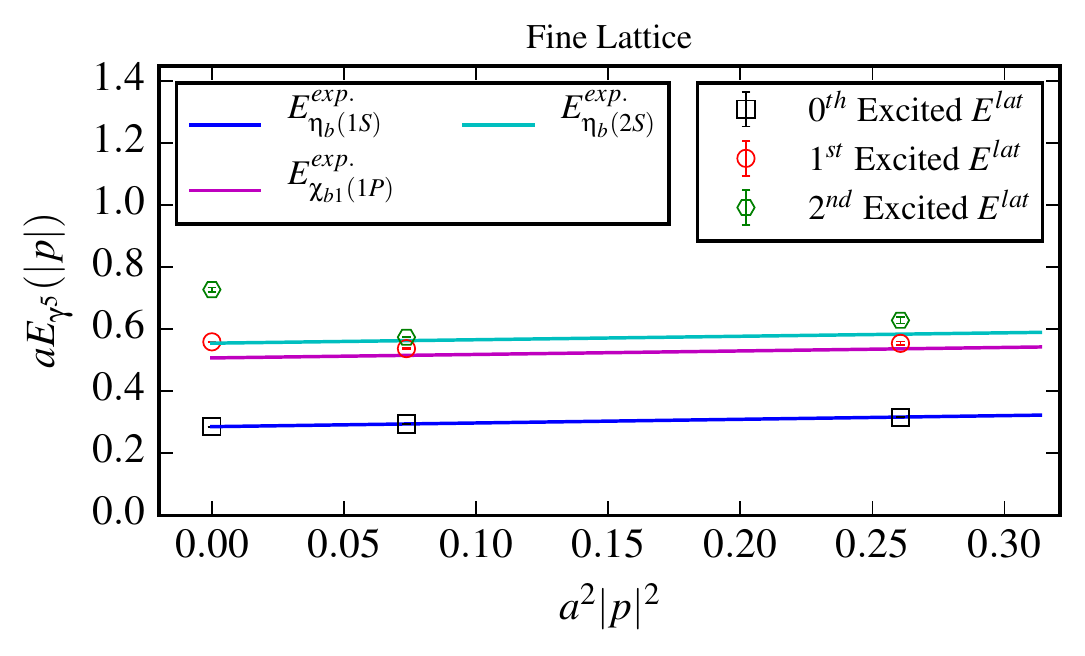}
  \caption{The first three energies extracted from the lattice NRQCD
    correlator data with the operator $\mathcal{O}^{\gamma^5}$ across multiple momenta. Statistical errors
    only. At nonzero momentum, the energy of the first excited state is
    lower than the energy of the first excited state at
    zero-momentum. This is a consequence of new states being present
    in the correlator data at nonzero momentum, as described in
    Section \ref{sec:EnergyEigenstates}. Thus, care must be taken not to misidentify states. $aE^{exp.}$ represents the energy of the states  according to a
nonrelativistic, rotational dispersion relation reconstructed using the experimental masses details of which can be found in the text.  }
  \label{fig:EnergyEta}
\end{figure}

\begin{figure}[t]
  \centering
  \includegraphics[width=0.49\textwidth]{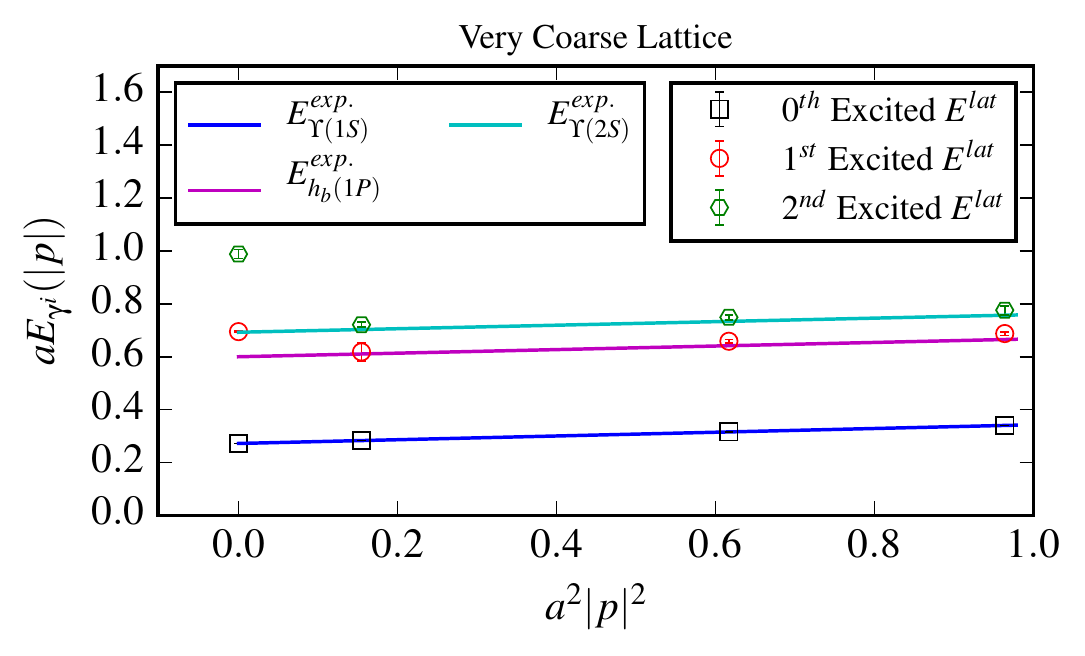}
  \includegraphics[width=0.49\textwidth]{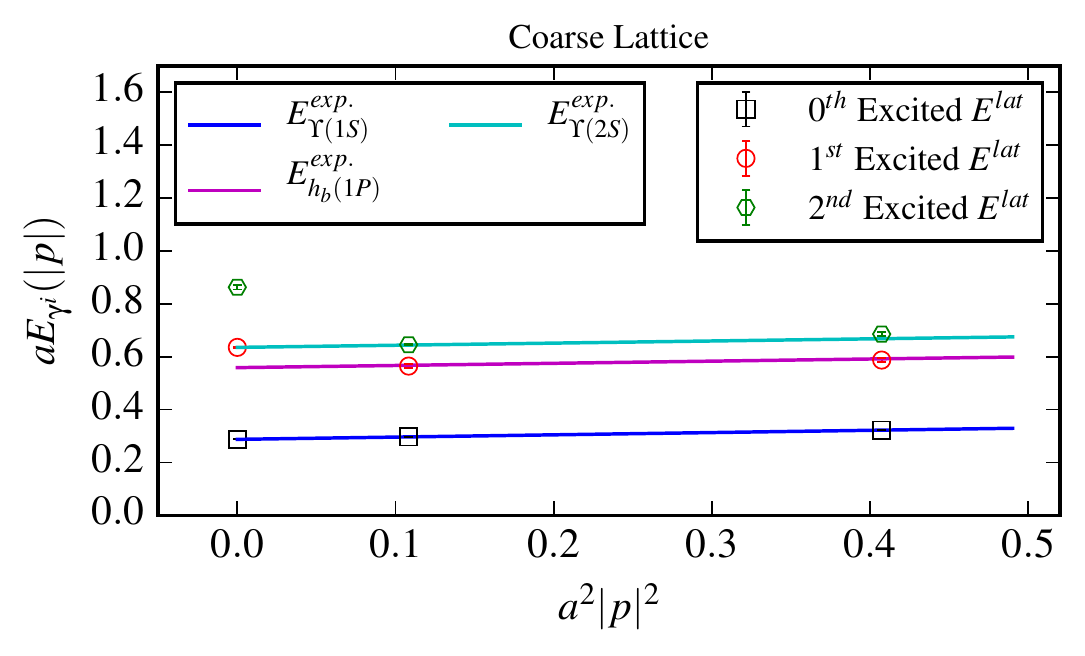}
  \includegraphics[width=0.49\textwidth]{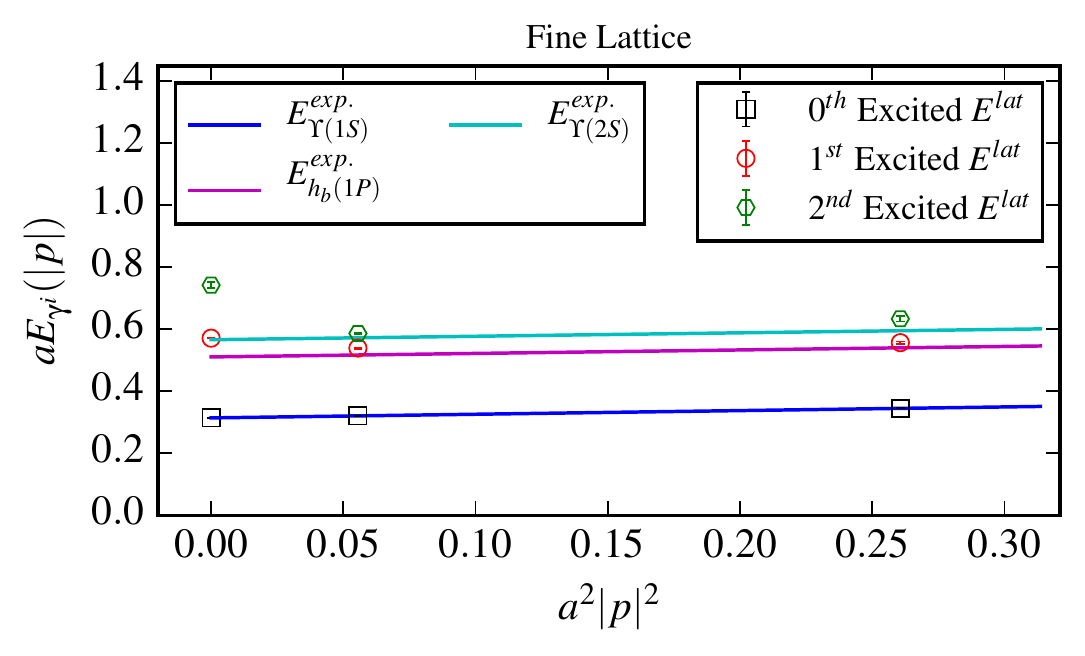}
  \caption{As in Figure \ref{fig:EnergyEta} but with  the operator $\mathcal{O}^{\gamma^i}$.}
  \label{fig:EnergyUps}
\end{figure}

Theoretically, particle states living in the Hilbert space are
classified in terms of invariant quantities within irreducible representations (irreps) of
the symmetry group of a theory. For our
calculation, two groups need to be considered: the Lorentz group and
the continuous rotational group in three dimensions. Appendix \ref{app:Classification} reviews the
construction of the irreps of both these groups at zero and nonzero
momentum. 

As is well known, the irreps of the Lorentz group at rest are
described by $|{p}^2=m^2;J^{PC},M\rangle$, where $J$, $M$ are the total
and third component of angular momentum respectively. $P$ is the
parity quantum number and for quarkonia $C$ is the charge
conjugation. The quantum
numbers $J^{PC}$ classify all particles seen in experiment to date \cite{PDG:2014}. 

However, the symmetry group
of  NRQCD is only the rotational group. At zero momentum, the states within such a theory are also
described by $|{\VEC{p}}={\VEC{0}};J^{PC},M\rangle$. At nonzero momentum, the situation is significantly different, and the irreps are described by $|{\VEC{p}} \ne {\VEC{0}};\lambda\rangle$, where $\lambda$ is an eigenvalue of the helicity operator $\hat{\lambda} = \hat{p}\cdot \hat{J}/E$.  This has important consequences for the energy spectrum
extracted from our lattice calculation (compare the zero and nonzero momentum lattice spectrum seen in Figures \ref{fig:EnergyEta}, \ref{fig:EnergyUps}) and therefore needs to be fully understood in order to have a reliable computation. 

At
rest the bilinear operators that we use in our calculation, listed in
Table \ref{tab:Operators} with $\Gamma = i\gamma^5, \gamma^i$, overlap onto definite $J^{PC} =
0^{-+}, 1^{--}$ energy eigenstates respectively in the infinite volume
continuum version of our theory (which is  rotationally invariant) \cite{Thomas:Helicity}. In Appendix \ref{app:Classification} (as in \cite{Thomas:Helicity}) it is shown that at nonzero momentum, $\mathcal{O}^{\gamma^5}({\VEC{p}})$ is a helicity operator which creates a definite $\lambda=0^-$ energy eigenstate, but
$\mathcal{O}^{\gamma^i}({\VEC{p}})$ creates an admixture of
$\lambda = 0^+, \pm 1$ eigenstates, where these $\lambda$ get
contributions from $J^P$ values as listed in the third column of Table
\ref{tab:Operators}. The $\pm$ superscript on the
$\lambda=0$  represents the eigenvalue $\tilde{\eta} \equiv P(-1)^J$
from  the $\hat{\Pi}$ symmetry  (a parity transformation followed by a rotation to bring the momentum direction back to the original direction) \cite{Thomas:Helicity}. 

In the correlator data from using $\mathcal{O}^{\gamma^5}({\VEC{p}}
\ne {\VEC{0}})$, guided by the experimental masses and this analysis, the lowest states in the spectrum should be $\eta_b(1S) (=0^{-+}), \chi_{b1}(1P)  (=1^{++}), \eta_b(2S) (=0^{-+})$, etc. whereas from using $\mathcal{O}^{\gamma^i}({\VEC{p}})$ the lowest states in the spectrum should be $\Upsilon(1S) (=1^{--}), h_{b}(1P)  (=1^{+-}), \Upsilon(2S) (=1^{--})$, etc. These are the $J^P$ states which we see in our lattice spectrum at nonzero
momentum.

The first three states extracted from the
spectrum with the operator $\mathcal{O}^{\gamma^5}$, $\mathcal{O}^{\gamma^i}$ are shown
in Figures \ref{fig:EnergyEta}, \ref{fig:EnergyUps} respectively. On
the same plot,  the solid lines represent the energy of the states  according to a
nonrelativistic, rotational dispersion relation reconstructed using the experimental masses, e.g., $aE(|{\VEC{p}}|) =
am^{\text{sim}} + |{\VEC{p}}|^2/2am^{\text{kin}}$, where $m^{\text{kin}}$ is the
kinetic mass which we set equal to the experimental mass, and
$m^{\text{sim}}$ is the static mass offset due to neglecting the mass term
in the NRQCD Hamiltonian. We find $am^{\text{sim}}$ in the correlator
data from the
$\mathcal{O}^{\gamma^5}$ operator by taking the ground state
lattice energy at zero momentum and finding the shift in the static mass
as the difference $a\Delta = am_{\eta_b(1S)}^{\text{exp.}} -
am_{\eta_b(1S)}^{\text{lat}}$. We then use this value of the
shift to
find  $am^{\text{exp,sim}}_{J^{PC}} =
am^{\text{exp.}}_{J^{PC}} - a\Delta$, to be used in the above dispersion
relation. We found the shift in the $\mathcal{O}^{\gamma^i}$
correlator data in the same way. 

The important
point to observe in these figures is that at nonzero momentum the
energy of the first excited state is
actually lower than the energy of the first excited state at
zero-momentum, opposite to what one would expect from a dispersion
relation. The reason is clear: at nonzero momentum energy eigenstates
have definite helicity, not definite $J^{P}$.  Therefore 
our correlator data gets contributions from the
$J^{P}$ states listed in Table \ref{tab:Operators}.

We conclude that, as Figures \ref{fig:EnergyEta} and \ref{fig:EnergyUps} show, one has to be careful in equating the states found in NRQCD at nonzero
momentum with continuum $J^{PC}$ quantum numbers and also in
extracting matrix elements involving a state inflight. However, here we only extract excited states at zero-momentum in order to avoid unnecesssary complications and to obtain high-precision
results, which can be muddled when extracting excited states in flight
due to the addition of extra states in the spectrum and their small
overlap factors as described in Appendix \ref{app:Classification}. After our analysis, we can then be sure that we have extracted the correct matrix element for the $\Upsilon(2S)\to\eta_b(1S)\gamma$ decay.

\subsection{Matrix Elements from Lattice QCD}
\label{sec:3pt}

\begin{figure}[t]
\begin{center}
\includegraphics[width=0.495\textwidth]{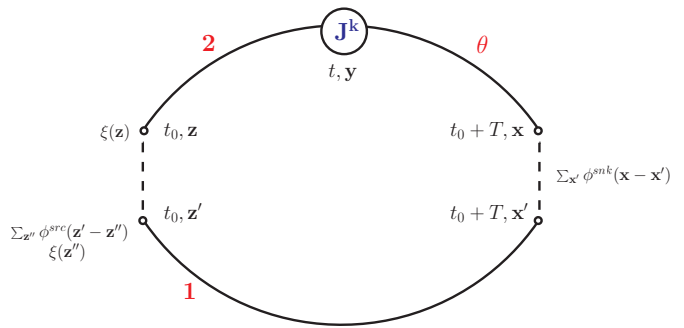}
\end{center}
\caption{ Setup for the three-point correlator calculation as described in
    Sec.~\ref{sec:3pt}. Propagator $1$ is the anti-quark and $\xi(x)$ is the random noise source as described in the text.}
\label{fig:ThreePoint}
\end{figure}

The simplest quantity which encodes information on a meson-to-meson decay matrix
element from within lattice QCD is the three-point correlator
\begin{align}
&{{C}}^{mn}_{\text{3pt}}(n_{src}, n_{snk}, {\VEC{p_f^{\theta} = -q^{\theta}}}
;t,T)  = \label{eqn:3ptCorr}  \\
&  \sum_{{\VEC{x,y}}} e^{-i {\VEC{x\cdot p^{\theta}}}}
\langle \mathcal{O}_f(n_{snk}; {\VEC{x}},T)
{{J^n}}({\VEC{q^{\theta}}}; {\VEC{y}},t )  \mathcal{O}_i^{m\dagger}(n_{src}; {\VEC{0}},0)  \rangle \nonumber
\end{align}
where $\mathcal{O}^{m}_{i}$, $\mathcal{O}_{f}$ are interpolating operators which create the
initial state with polarisation $m$ and final state respectively,
${{J^n}}({\VEC{q^{\theta}}}; {\VEC{y}},t ) = \psi^{\dagger} \Gamma^n({\VEC{q^{\theta}}}; {\VEC{y}},t )  \psi$ is the current which
induces the transition with $n$ labelling the polarisation of the photon, and the twisted momenta are described in Sec.~\ref{sec:Mom}. The three-point correlator is visualised as in Figure \ref{fig:ThreePoint} where the three points in lattice units correspond to: the source point
of the initial particle at time $t_0$ (equal to zero in (\ref{eqn:3ptCorr})); the position and time of the current
causing the transition at (${\VEC{y}}$, $t$); and the position and time of the
final state at (${\VEC{x}}$, $T$). After performing Wick contractions on
the three-point correlator the connected contribution, written
in terms of NRQCD propagators as discussed in Section \ref{sec:NRQCD},
is
\begin{widetext}
\begin{align}
{{C}}^{mn}_{\text{3pt}}(n_{src}, n_{snk}, {\VEC{p_f^{\theta} = -q^{\theta}}}
;t,T) & = -\sum_{{\VEC{x,y}}} e^{-i {\VEC{x\cdot p^{\theta}}}}
\text{Tr} \Big[ \Gamma^{m}_i G_{\chi}({{0}}|{{x}} ) \Gamma_f \tilde{G}^{\theta}_{\psi}({{x}}|{{y}}) {{\Gamma^{n}}}({\VEC{q^{\theta}}}; {{y}}) G_{\psi}({{y}}|{0})  \Big] \label{eqn:3ptWick}
\end{align}
\end{widetext}
where the twisted propagator $\tilde{G}^{\theta}(x|y)$ is defined in
Appendix \ref{app:TwistedCorrelators}. Direct computation of the propagator $G(x|y)$ is
unnecessarily expensive as we can use the sequential source technique (SST)
\cite{SequentialSource, Dudek:CharmRad}  to yield the desired
propagator, which only requires one further evolution.
There are two ways to package the $G(x|y)$ propagator in the
three-point correlator when using the SST. The first is called the fixed
current method, which requires the insertion time $t$ to be fixed and for propagator $2$ in Figure \ref{fig:ThreePoint} to be used as a source for propagator $\theta$. However, this method does not scale well and is undesirably expensive for relativistic quark formalisms.

The second approach is called the fixed sink method. In this approach,
one fixes the sink time $T$ and factorises (\ref{eqn:3ptWick}) as
\begin{align}
&{{C}}^{mn}_{\text{3pt}}(n_{src}, n_{snk}, {\VEC{p_f^{\theta} = -q^{\theta}}}
;t,T) \nonumber \\ 
& \hspace{0.2cm} = -\sum_{{\VEC{y}}} e^{-i {\VEC{y\cdot \theta}}} \text{Tr} \Big[ \Gamma^m_i H^{\theta \dagger}(y|0) {{\Gamma}^n}({\VEC{q^{\theta}}}; {{y}}) G_{\psi}({{y}}|{0})  \Big] \label{eqn:3ptWickFixedSink}
\end{align}
with 
\begin{align}
H^{\theta}(y|0) & = \sum_{{\VEC{x}}} e^{i {\VEC{x\cdot p}}}
{G}^{\theta}_{\psi}({{y}}|{{x}} ) \Gamma^{\dagger}_f
{G}_{\psi}({{x}}|{{0}}) \nonumber
\end{align}
where we have written $H(y|0)$ in terms of the twisted propagator that
satisfies periodic boundary conditions 
and used the fact that $\Gamma_f$ commutes with the
exponential as described in Appendix \ref{app:TwistedCorrelators}. We
have also used the NRQCD
$\gamma^5$-hermicity conditions from Sec.~\ref{sec:NRQCD}, and used
$G^{\dagger}_{\psi}(x|y) = - G_{\psi}(y|x)$  because $G(x|y) =
\langle \psi(x)\psi^{\dagger}(y) \rangle$. We can obtain
$H^{\theta}(y|0)$ by using the  twisted evolution
equations with the source $e^{i {\VEC{x\cdot p}}}
\Gamma^{\dagger}_f {G}_{\psi}({{x}}|{{0}})$. 

Clearly, the two methods should give the same correlator data as they only
differ in how $G(x|y)$ is packaged. We have checked this
numerically and found it to be true on any given configuration up to machine
precision. As the fixed sink method is more cost effective, this method
was used for the calculation. Our program structure can be
visualised in Figure \ref{fig:ThreePoint}. Propagator $1$ is generated
with a smeared, random wall source at time $t_0$ and propagated to
time $T$ where the sink smearing is applied. $H^{\theta}(y|0)$ is
found by using the source $e^{i {\VEC{x\cdot p}}}
\Gamma^{\dagger}_f{G}^{1}_{\psi}({{x}}|{{0}})$ and
evolving backwards in time using the twisted configurations to a time
$0\le t\le T$. Propagator $2$ is made from the same random wall as
$1$. We then combine propagator $2$, $H^{\theta}(y|0)$ and the current
as in (\ref{eqn:3ptWickFixedSink}) to obtain the three-point
correlator. We use the same $16$ time sources as in the two-point
correlator and prior to fitting, all data is translated to a common
$t_0=0$. 

The three-point correlator (\ref{eqn:3ptCorr}) can be related to matrix
elements of the current by inserting a complete set of states
\cite{Dudek:CharmRad}. By doing so, and using the rotational
parameterisation of the overlaps as described in Appendix
\ref{app:Classification}, $C^{mn}_{\text{3pt}}$ is seen to be
anti-symmetric. We
average over the six nonzero contributions using an isotropic momentum  as
\begin{align}
C^V_{\text{3pt}} & = \frac{1}{6} \sum_{l=1}^{3} \epsilon_{lmn}
C^{nm}_{\text{3pt}} \,. \label{eqn:3ptCorrAvg}
\end{align}

In addition,  inserting the complete set of states also leads to the functional form of the fitting function
\begin{align}
&C^{V}_{\text{3pt}}(n_{src}, n_{snk}, {\theta} ;t,T) \nonumber \\ 
&\hspace{0.5cm} = \sum_{{{i,
      f}}} a(n_{snk},i) V^{\text{fit}}_{if} b^*(n_{src},f) e^{-E_it}e^{-E_f(T-t)}  
\label{eqn:3ptCorrFit}
\end{align}
where $a(n_{snk},i)$ and $b(n_{src},f)$ are amplitudes from the
two-point fitting function in (\ref{eqn:2ptCorrFit}). The two-point and three-point correlators can be simultaneously fit to
(\ref{eqn:2ptCorrFit}) and (\ref{eqn:3ptCorrFit}) respectively using
multi-exponential chained \cite{ChainedFit}, marginalised \cite{MarjFit} Bayesian fitting. Chained,
marginalised fitting has been shown to significantly decrease the
fitting time and produce reliable, precise and accurate results if the data is
in the limit of high statistics (Gaussianly distributed) \cite{FastFits}. We check that results are compatible from both with and without chained, marginalised fits on a
subset of the data. We use a prior of $0.1(0.2)$ for all
$V^{\text{fit}}_{if}$  and the same priors for the amplitudes and energies as in the two-point fits described in Sec.~\ref{sec:2pt}. For each current, we obtain data
for fixed $T=9,12,15$ and the same $3\times 3$ matrix of smearings as in the
two-point correlators.  This allows accurate 
extractions of the matrix element as it includes excited state
contributions. 

The use of a singular value decomposition stabilises
the fit and is standard practice in the literature \cite{ChainedFit}. In our Bayesian fit, this is performed by setting a tolerance and replacing all eigenvalues of the correlation matrix smaller than this tolerance times the maximum
eigenvalue to this value \cite{ChainedFit}. By doing so, this leads to larger errors in
the fit results and so is a conservative step. We use a tolerance of $10^{-4}$. 

The matrix element for the $\Upsilon(2S)\to\eta_b(1S)\gamma$ decay will be
proportional to $V_{21}^{\Upsilon\eta_b}$. By equating the fitting functions to
their continuum correlator counterparts with conventional relativistic
normalisation, parameterising our overlaps
using rotational invariance with the initial particle at rest, we find
\begin{align}
V_{21}^{\Upsilon\eta_b}(q^2) & = \frac{m_{\Upsilon(2S)} +
  m_{\eta_b(1S)}}{m_{\Upsilon(2S)}\theta_i}
\sqrt{m_{\Upsilon(2S)}E_{\eta_b}} V^{\text{fit}}_{21} \label{eqn:FormFactorFit}
\end{align}
where ${\VEC{\theta}}$ is the twisted momentum described in Sec.~\ref{sec:Mom}. Since the static masses obtained from an NRQCD calculation are shifted, as
explained previously, we extract $V^{\Upsilon\eta_b}_{21}(q^2)$ from
$V^{\text{fit}}_{21}$ using the same experimental masses as in Sec.~\ref{sec:DecayRate}. A nonrelativistic dispersion relation was used to find $E_{\eta_b(1S)}$, which is appropriate as shown in Figure
\ref{fig:EnergyEta}.


\section{M1 Radiative Decay Currents}
\label{sec:Currents}

In order to compute the form factor ${V}^{\Upsilon\eta_b}_{nm}(q^2)$, we need
to choose currents which will induce a hindered M$1$ radiative
decay. Within a nonrelativistic framework, it is a standard result in the
literature \cite{Feinberg:M1, Sucher:M1, Kang:1978} that the leading order contribution to the matrix
element is suppressed due
to the orthogonality of the radial wavefunctions and relativistic
corrections are necessary. This suppression
introduces a sensitivity to a range of effects that we must test and
quantify in order to perform an accurate calculation. The first of
these effects is the fact that next-to-leading order current
contributions are appreciable and  we need to include
them. 

As we are using NRQCD to simulate the $b$-quark, choosing the currents
from a NRQCD and non-relativistic quantum electrodynamics (NRQED) effective field theory is most appropriate. This
effective field theory can be found straightforwardly by extending the
$SU(3)$ Lie algebra of NRQCD to a $SU(3) \times U(1) $ Lie algebra to
produce NRQCD $+$ NRQED \cite{Brambilla:M1}.  Then, in principle, one
could discretise the $SU(3) \times U(1)$ theory and choose appropriate
currents from the resulting operators. However, this introduces
complications, e.g.\ the $U(1)$ magnetic field only decouples from the
$SU(3)$ chromomagnetic field to leading order in the lattice spacing,
resulting in lattice artefact currents which are not present in the
continuum.  Calculating such currents would require more computational
resources and make the computation of the matching coefficients more
difficult.

Instead, we are free to choose the currents from the continuum NRQCD
$+$ NRQED theory and renormalise these. It is important to understand the power counting in the NRQCD $+$ NRQED effective field theory in order to choose our currents appropriately. Given that NRQCD $+$ NRQED is a $SU(3) \times U(1)$ effective field theory, it has two expansion parameters. For NRQCD, we have the standard expansion
parameter $v$, where $v^2 \sim 0.1$ for bottomonium. The only scale
available for the on-shell emitted photon is the photon's energy $|
\vec{q}_{\gamma}| \sim 0.6 $ GeV. Since the photon's energy is the
difference between the masses of two heavy S-wave quarkonia, it is of
the order $|\vec{q}_{\gamma}| \sim mv^2 \sim 0.4$ GeV. Thus we can
expand our effective field theory in terms of
$v$ only.

We summarise the power counting as
\begin{itemize}
\item $A_{QED} \sim |\vec{q}_{\gamma}|$.
\item $B_{QED}$, $E_{QED} \sim |\vec{q}_{\gamma}|^2$. 
\item The standard QCD power counting rules for the QCD fields.
\item The knowledge that when a derivative acts on the photon field, it gives a
factor of $|\vec{q}_{\gamma}|$ and when acting on the quark field a
factor of $p_q \sim mv$ (as the valence quark knows nothing of the
 photon momentum in the initial quarkonium rest frame).
\end{itemize}

Ordering the operators that induce a M$1$ (spin-flip) transition from
NRQCD $+$ NRQED, we find (to next-to-leading order for our decay and
borrowing notation from \cite{KinNio96})
\begin{align}
 O_F  &= \omega_F \frac{ e e_b }{2m_b} \psi^{\dagger}  {\VEC{{\sigma}
\cdot {B}_{QED}}} \psi 
\nonumber \\
 O_{W1} &= \omega_{W1} \frac{ e e_b }{8m_b^3}  \psi^{\dagger} \{
{\VEC{ {D}}}^2 , {\VEC{{\sigma}
\cdot {B}_{QED}}} \} \psi
\nonumber \\
 O_{S} &= \omega_S \frac{i e e_b }{8m_b^2} \psi^{\dagger} {\VEC{ \sigma \cdot ( {D}
\times {E}_{QED} - {E}_{QED} \times {D} )}} \psi  \nonumber \\
 O_{S2} &= \omega_{S2} \frac{i 3 e e_b }{64 m_b^4} \times \nonumber \\ 
& \hspace{0.0cm} \psi^{\dagger} \{
{\VEC{ {D}}}^2, {\VEC{ {\sigma} \cdot ( {D}
\times {E}_{QED} - {E}_{QED} \times {D} )}} \} \psi
\nonumber \\
 O_{tot} &= O_F + O_{W1} + O_{S} + O_{S2} \label{eqn:CurrentsKN}
\end{align}
Here,  $i\vec{D} = i\vec{\nabla} + g
\vec{A}^a_{QCD}T^a$ are all pure QCD covariant derivatives, fields marked
QED (QCD) are the QED (QCD) fields and $\omega_i$ are the matching coefficients needed to reproduce full QCD$+$QED from our effective theory. Using the power counting rules above, we find $ O_F  \sim v^4$, $O_{W1} \sim v^6$,  $O_{S} \sim v^5$ and  $O_{S2} \sim v^7$. We can then factor out the
photon and electric charge in order to derive the currents
$J_k({\VEC{q^{\theta}}}; {\VEC{y}},t)$ which give the
decomposition of the matrix element in (\ref{eqn:MatrixElement}). For
example, the operator $O_F$ gives rise to the current
\begin{align}
& { J_F^k}  = -\omega_F \frac{ 1 }{2m_b} \psi^{\dagger} {\VEC{ ( {\sigma}
\times }} {i{\VEC{ q}}})^k e^{-i {\VEC{ q \cdot x}} } \psi \,.
\nonumber
\end{align}

We then write all currents as $J^k({\VEC{q^{\theta}}}; {\VEC{y}},t)
= \psi^{\dagger}\Gamma_k({\VEC{q^{\theta}}}; {\VEC{y}},t) \psi$
so that $\Gamma_k({\VEC{q^{\theta}}}; {\VEC{y}},t)$ will be
what enters the three-point correlator as in (\ref{eqn:3ptWickFixedSink}). We use the
terminology that the form factor coming from the current $J_F$ is
called $V^{\Upsilon\eta_b}_{21}|_F
= \omega_F \tilde{V}^{\Upsilon\eta_b}_{21}|_F$, where the tilde implies we have factored off the
matching coefficient from the form
factor in the numerical calculation  and this should be applied later
in the analysis. Similar notation is used for the other currents and we refer to $\tilde{V}^{\Upsilon\eta_b}_{21}|_i$ as unrenormalised form factors. The
final form factor is $V^{\Upsilon\eta_b}_{21} = \sum_i \omega_i\tilde{V}^{\Upsilon\eta_b}_{21}|_i$. 

It should be noted that there are other currents (suppressed by $v$ or $\alpha_s$)  that contribute to this decay and which might be of interest, notably, the
 QCD analogues of the $O_{W1}, O_{S}$, operators  
arising from choosing the electric (magnetic) fields in
(\ref{eqn:CurrentsKN}) to be gluon
fields and the photon coming from the full $SU(3) \times U(1)$
covariant derivative. Other
 operators are those which only occur at loop level in
the full QCD $+$ QED theory. These can be written as
\begin{widetext}
\begin{align}
O_{W1QCD} &= -\omega_{W1QCD} \frac{ i e e_b }{8m_b^3}  \psi^{\dagger} \{
{\VEC{ {A}_{QED} \cdot {D} + {D} \cdot {A}_{QED} , {\sigma}
\cdot g {B}_{QCD}}} \} \psi \nonumber \\
 O_{SQCD} &= \omega_{SQCD} \frac{e e_b }{8m_b^2} \psi^{\dagger}
{\VEC{ {\sigma} \cdot ( {A}_{QED} \times g {E}_{QCD} - g {E}_{QCD} \times {A}_{QED} )}} \psi  \nonumber \\
 O_{W2} &= \omega_{W2} \frac{ e e_b }{4m_b^3}  \psi^{\dagger}
{D}^i {\VEC{{\sigma} \cdot {B}_{QED}}} {D}^i \psi
\nonumber \\
 O_{p'p} &= \omega_{p'p} \frac{ e e_b }{8m_b^3}  \psi^{\dagger}
 {\VEC{ {\sigma} \cdot {D} }} {\VEC{ {B}_{QED} \cdot {D}}} +
 {\VEC{ {D}
 \cdot {B}_{QED}}} {\VEC{{\sigma} \cdot {D}}}  \psi \,.
\label{eqn:AdditionalCurrentsKN}
\end{align}
\end{widetext}
When attempting power counting on the QCD operators above, it is helpful
to draw the Feynman diagram that such an operator would
produce. Essentially, we need to contract the gluon field with another,
producing another factor of $g v^3$ at least \cite{Lepage:ImprovedNRQCD}. 
Consequently these operators
are expected to be of order $\alpha_s v^8$ at most. We confirm numerically that the form factors from these QCD operators are suppressed as expected and they are negligible within the errors of our final results. Since $\omega_{W2}, \omega_{p'p}$ occur only at loop level they are suppressed by $\mathcal{O}(\alpha_s)$ relative to $O_{W1}$. We will introduce a systematic error for neglected currents in the final analysis. 

\subsection{Matching Coefficients for the Currents}
\label{sec:MatchingCoeff}

The matching coefficients, $\omega_i$,  appearing in the operators in (\ref{eqn:CurrentsKN}) are needed to take into account the high-energy UV modes from processes in the full theory but not present in our effective field theory. They have the
expansion $1 + \omega_i^{(1)}\alpha_s + \mathcal{O}(\alpha_s^2)$. Here
we compute the one loop correction to the coefficient $\omega_F$ from the leading order current. Following this, we estimate the
errors from neglecting corrections that we do not calculate.

Our calculation of the one-loop coefficient $\omega_F^{(1)}$ is very similar to the computation of the one-loop
correction of $c_4$ in \cite{Hammant:2013}. Following that analysis, by matching the current
from NRQCD + NRQED to continuum QCD+QED, we find
\begin{align}
\omega_F^{(1)} & =  b^{(1)}_{\sigma,QED}- Z_m^{NR,(1)} - Z_m^{tad,(1)} \nonumber \\
& \hspace{1.5cm} -Z_2^{NR,(1)} - Z_{\sigma,QED}^{NR,(1)}  \label{eqn:OmegaF}
\end{align}
where $b^{(1)}_{\sigma,QED} = {C_F}/{2\pi}$ is the coefficient of the first order
correction to the quark's magnetic moment, computed analytically in continuum QCD 
following standard techniques. As $b_{\sigma,QED}$ is UV
finite, this allows us to directly equate results obtained on the
lattice to those obtained in the continuum, since the difference between
the schemes for UV regulation is then irrelevant. In the general matching procedure the continuum and lattice IR divergences cancel in the computation of the radiative correction; here, because of the standard Ward Identity, the continuum and lattice contributions to $\omega_F^{(1)}$ are separately finite.

$Z^{NR}_m$, $Z^{NR}_2$, $Z^{NR}_{\sigma,QED}$ are the renormalisation factors
of the bare quark mass, the
wavefunction and the current $J_F$ from (\ref{eqn:CurrentsKN}). These are
calculated in lattice NRQCD. We automatically generate the Feynman rules for a specific NRQCD
action (along with the Symanzik-improved gluonic action) using the
HiPPy package, then compute the numerical evaluation of
these diagrams using the HPsrc package \cite{HIPPY, HIPPY2}. We use the full
$v^4$ NRQCD Hamiltonian with spin dependent $v^6$ pieces  as defined in
(\ref{eqn:LNRQCDv6}). Computation of $Z^{NR,(1)}_m$ and $Z^{NR,(1)}_2$
is identical to \cite{Hammant:2013}.  $Z^{NR,(1)}_m$ will get
contributions from mean-field corrections which we denote as $Z_m^{tad,(1)}$. We use the Landau mean link $u_0^{(2)} = 0.750$
\cite{Nobes:2001}. For
the action that we use, the
tadpole correction is \cite{Hammant:2013}
\begin{align}
 Z_m^{tad} & = - \left( \frac{2}{3} + \frac{3}{(am_b)^2}  \right)
 u_0^{(2)} \,. \label{eqn:ZmTad}
\end{align}

\begin{figure}[t]
  \centering
  \includegraphics[width=0.20\textwidth]{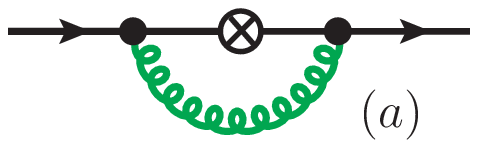}
  \includegraphics[width=0.20\textwidth]{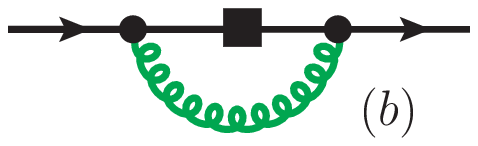}
  \includegraphics[width=0.20\textwidth]{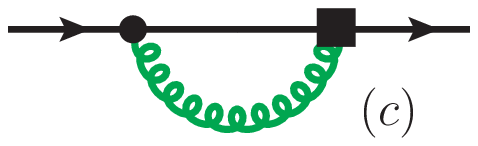}
  \includegraphics[width=0.20\textwidth]{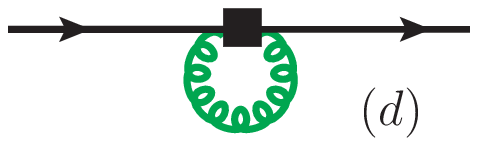}
  \caption{Classes of one-loop diagrams which contribute to
    $Z^{(1)}_{\sigma,QED}$ as described in the text. The cross inside a
    circle represents the $J_F$ current obtained from
    (\ref{eqn:CurrentsKN}), while the solid box represents the
    higher order currents from (\ref{eqn:CurrentsKN}) and the exchange of a gluon is denoted by a curly line.  }
  \label{fig:FeynmanDiagrams}
\end{figure}

The NRQCD diagrams contributing to $Z_{\sigma,QED}^{NR,(1)}$ are shown in
Figure \ref{fig:FeynmanDiagrams}. Since we do not actually include the QED
field in our calculation, there are no tadpole factors from this term. Note
that Fig.~\ref{fig:FeynmanDiagrams}(a) is generated by the current coming from
$ \psi^{\dagger}\VEC{\sigma\cdot B_{QED}}\psi /2am_b$ being inserted at the
vertex, and Figs.~\ref{fig:FeynmanDiagrams}(b), \ref{fig:FeynmanDiagrams}(c)
and \ref{fig:FeynmanDiagrams}(d) arise from mixing effects from the
higher order currents (that we include in the calculation of the decay rate)
from (\ref{eqn:CurrentsKN}). Computation of the Feynman diagrams shown in
Figs.~\ref{fig:FeynmanDiagrams}(b),~\ref{fig:FeynmanDiagrams}(c) and \ref{fig:FeynmanDiagrams}(d) is more
involved than that of Fig.~\ref{fig:FeynmanDiagrams}(a), so 
they are not included in
this calculation, but we plan on computing them in future work.  For now, we will
introduce a systematic error from neglecting these contributions.

A breakdown of the numerical values of the various terms that enter
$\omega_F^{(1)}$ for the masses that we use in
this calculation is shown in Table \ref{tab:OmegaF}. $\omega_F^{(1)}$ was computed for a range of masses
(neglecting the mixing down) and we give these values in Table \ref{tab:OmegaFAll}. 

\begin{table}[t]
  \caption{ Breakdown of the different terms that go into
    $\omega_F^{(1)}$.  The $\alpha_s(q^* = \pi/a)$ values are taken from
    Table \ref{tab:NRQCDParams}. }
\label{tab:OmegaF}
\begin{center}
  \begin{tabular}{|c | c | c| c | }
    \hline 
    $am_b$ & $1.90$ & $2.70$ & $3.30$ \\ \hline\hline
    $Z_m^{(1)} +Z_2^{(1)} +Z_{\sigma,QED}^{(1)}$ & $1.2961(5)$ &
    $0.9061(4)$ &  $0.7585(6)$ \\ 
    $ Z_m^{tad}$ & $-1.1233$ & $-0.8086$ & $-0.7066$  \\
    $\omega_F^{(1)}$ & $0.0394(5)$ & $0.1148(4)$ & $0.1603(6)$ \\
    $\alpha_s({\pi}/{a})\omega_F^{(1)}$ & $0.0089(1)$ & $0.0293(1)$
    & $0.0441(2)$ \\
 \hline 
  \end{tabular}
\end{center}
\end{table}

\begin{table}[t]
  \caption{ Values of $\omega_F^{(1)}$ at various $am_b$ values.  }
\label{tab:OmegaFAll}
\begin{center}
  \begin{tabular}{l c c c }
    \hline \hline
    $am_b$ & $1.1$ & $1.5$ & $2.1$  \\ \hline
    $\omega_F^{(1)}$ & $-0.211(2)$ & $-0.030(1)$ & $0.0626(9)$   \\
    \hline \hline
    $am_b$ & $2.4$ & $4.0$ & $4.6$ \\ \hline
    $\omega_F^{(1)}$ &  $0.0918(7)$  & $0.2039(4)$ & $0.2372(5)$ \\
 \hline \hline
  \end{tabular}
\end{center}
\end{table}

\begin{figure}[t]
  \centering
  \includegraphics[width=0.49\textwidth]{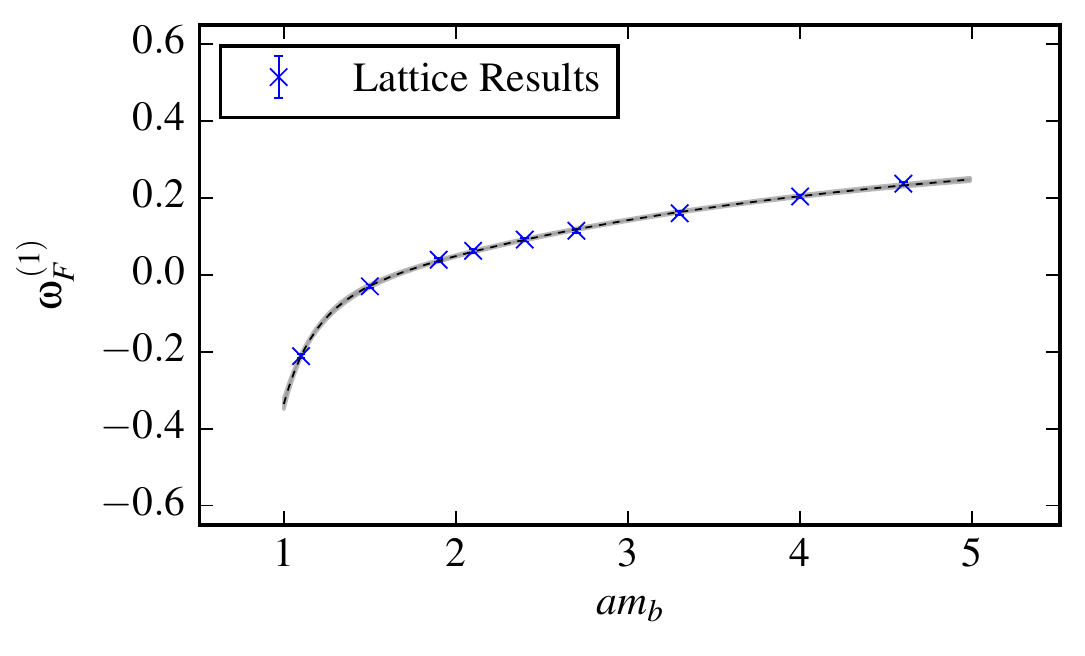}
  \caption{The values $\omega_F^{(1)}$ with a smooth interpolating curve as described in the text. }
  \label{fig:OmegaFFit}
\end{figure}

We show the  values of $\omega_F^{(1)}$ with a smooth interpolating curve in Figure \ref{fig:OmegaFFit}. This interpolating curve was chosen to be a polynomial in
$1/am_b$ in order to reproduce the static limit as
$m_b\to\infty$. To fit these values easily we increased the errors on the points
returned by HPsrc to $1\%$. We use a Bayesian fit to all points in Figure \ref{fig:OmegaFFit} against a polynomial in
$1/am_b$. We found the smallest $\chi^2/dof (dof) = 0.7 (9)$ and largest Gaussian Bayes Factor \cite{Lepage:Code} when including all terms in the polynomial up to and including the quartic term. We used a prior for the constant piece as the polynomial of $0.4(2)$ and priors for the coefficients of the $1/(am_b)^n$ pieces of $0(1)$.

\subsection{Systematic Error from Current Matching Coefficients}
\label{sec:ErrorMatchingCoeff}

We need to include a systematic error from not knowing the matching
coefficients in the currents to infinite precision. There are two distinct
types of errors in this case: the first is from neglecting the $\mathcal{O}(\alpha_s^2)$ corrections
in $\omega_F$ and the $\mathcal{O}(\alpha_s)$ corrections to the
matching coefficients of the other currents;  the second is from neglecting the mixing
down effects on the values of $\omega_F^{(1)}$ used in the calculation. We will estimate each of
these in turn. 

To estimate the effect of neglecting the higher order corrections that we
have not calculated, it is helpful to compare our result to the pure
NRQED calculation of \cite{KinNio96}. There, the authors find that the
continuum QED contribution to their $\omega_F^{(1)}$ is the anomalous
magnetic moment of the electron $\alpha/2\pi$, while for us it is the
anomalous magnetic moment of the quark $C_F \alpha_s/2\pi$. For the
NRQED contribution, they find no IR log nor a constant piece and in their continuum approach the UV power law divergences may be omitted. Although we find no IR
log in our data, we cannot neglect the UV power law divergence associated
with the momentum cutoff. This shows up as a polynomial in $1/am_b$ as
mentioned above. We observe that this lattice artefact contribution gives a negative contribution to the
continuum value, as shown in
Table \ref{tab:OmegaF}, and for the $am_b$ range that we are interested in
$|\alpha_s \omega_F^{(1)}| <  C_F \alpha_s/2\pi$. As we are observing
similar behaviour over this mass range as the pure NRQED calculation,
we can use that calculation to estimate the error conservatively. 

As shown in \cite{KinNio96} and confirmed by the small values of our
numerical data, the matching coefficients can actually be expanded in
$\alpha_s/\pi$. In principle, the second order coefficient of $\omega_F$ could be $\mathcal{O}(1)$, and then this contribution could be
$\mathcal{O}(\alpha_s^2/\pi^2)$. As such, we allow for an additive systematic
error (assumed to be correlated across all ensembles) of $1 \pm \alpha_s^2/\pi^2$ from not knowing higher order
contributions to $\omega_F$.


We have not included the $\mathcal{O}(\alpha_s/\pi)$ contributions to the other
matching coefficients in (\ref{eqn:CurrentsKN}), namely $\omega_S$,
$\omega_{W1}$ and  $\omega_{S2}$. A difficult calculation would be
necessary to determine them. Again, we use the
equivalent parameters from the pure NRQED calculation \cite{KinNio96}
to estimate the systematic error. The pure NRQED equivalent of $\omega_{W1}$ has log
contributions in its first order coefficient and so we allow for an
additive correlated systematic error of $1\pm\alpha_s/\pi$ to the tree 
level value. We allow the same error on $\omega_{S2}$. 

The one loop correction of the pure NRQED equivalent of $\omega_S$ was
found to be $ 2\omega_F^{(1)} = \alpha / \pi$. As such, we allow for an additive correlated
systematic error on our $\omega_S$ of $1\pm  C_F\alpha_s/\pi$, to
compensate for using the tree level value in the calculation of the
decay rate. This is a conservative estimate as we see above that the
lattice artefacts actually subtract away some of this contribution
over the mass range we are interested in. 

The mixing down effects from diagrams (b), (c) and (d) in Figure
\ref{fig:FeynmanDiagrams} are difficult to estimate since each graph
by itself can be IR divergent but $\omega_F^{(1)}$ is IR finite. We allow an uncertainty of $30\%$ in the one-loop coefficient (correlated across all lattice spacings) from neglecting the mixing down. There is no substitute for the actual calculation though, and we intend to do this in the future. 


\section{Results For The $\Upsilon(2S)\to\eta_b(1S)\gamma$ Decay}
\label{sec:DecayNRQCD}

\begin{table}[t]
  \caption{ Values of the unrenormalised form factors
    $\tilde{V}^{\Upsilon\eta_b}_{21}|_i$, as described in Section
    \ref{sec:DecayNRQCD}, from the lattice NRQCD data on the ensemble labeled
    set $1$ in Table \ref{tab:GluonEnsembles}. We also give elements of
    the correlation matrix.
    A value of $a^2q^2 = 0.0034(21)$ was found from the data. } 
\label{tab:ViSet1}
\begin{center}
  \begin{tabular}{ l | r | r r r }
    \hline \hline
     $p$ & Value & $C(p , \tilde{V}|_{F} )$ & $C(p , \tilde{V}|_{W1} )$  &  $C(p,\tilde{V}|_{S} )$ \\ \hline
    $\tilde{V}^{\Upsilon\eta_b}_{21}|_{F}$  & $0.1818(42)$    & & & \\ 
    $\tilde{V}^{\Upsilon\eta_b}_{21}|_{W1}$ & $ -0.0594(12)$  & $-0.4010$ &           &\\  
    $\tilde{V}^{\Upsilon\eta_b}_{21}|_{S}$  & $ -0.0339(17)$  & $-0.2932$ & $0.1261$  &\\
    $\tilde{V}^{\Upsilon\eta_b}_{21}|_{S1}$ & $ -0.0037(3)$ & $-0.0624$ & $0.3488$ & $-0.2678$ \\ 
    \hline \hline
  \end{tabular}
\end{center}
\end{table}

\begin{table}[t]
  \caption{ Values and correlation matrix elements of the
    $\tilde{V}^{\Upsilon\eta_b}_{21}|_i$ from the ensemble labeled set $2$ in Table \ref{tab:GluonEnsembles}. A value of $a^2q^2 = 0.00338(92)$ was found from the data.} 
\label{tab:ViSet2}
\begin{center}
  \begin{tabular}{ l | r | r r r }
    \hline \hline
    $p$ & Value & $C(p , \tilde{V}|_{F} )$ & $C(p , \tilde{V}|_{W1} )$  &  $C(p,\tilde{V}|_{S} )$ \\ \hline
    $\tilde{V}^{\Upsilon\eta_b}_{21}|_{F}$  & $0.1765(22)$     & & & \\ 
    $\tilde{V}^{\Upsilon\eta_b}_{21}|_{W1}$ & $ -0.0593(7)$  & $-0.5298$ &           &\\  
    $\tilde{V}^{\Upsilon\eta_b}_{21}|_{S}$  & $ -0.0293(8)$  & $-0.3803$ & $0.3065$  &\\
    $\tilde{V}^{\Upsilon\eta_b}_{21}|_{S1}$ & $ -0.0045(2)$  & $-0.0134$ & $0.3962$ & $-0.2264$ \\ 
    \hline \hline
  \end{tabular}
\end{center}
\end{table}

\begin{table}[t]
  \caption{ Values and correlation matrix elements of the
    $\tilde{V}^{\Upsilon\eta_b}_{21}|_i$ from set $3$ in Table \ref{tab:GluonEnsembles}. A value of $a^2q^2 = 0.0007(12)$ was found from the data.} 
\label{tab:ViSet3}
\begin{center}
  \begin{tabular}{ l | r | r r r }
    \hline \hline
    $p$ & Value & $C(p , \tilde{V}|_{F} )$ & $C(p , \tilde{V}|_{W1} )$  &  $C(p,\tilde{V}|_{S} )$ \\ \hline
    $\tilde{V}^{\Upsilon\eta_b}_{21}|_{F}$  & $0.1720(36)$     & & & \\ 
    $\tilde{V}^{\Upsilon\eta_b}_{21}|_{W1}$ & $ -0.0577(10)$   & $-0.2634$ &           &\\  
    $\tilde{V}^{\Upsilon\eta_b}_{21}|_{S}$  & $ -0.0309(12)$   & $-0.1887$ & $0.2733$  &\\
    $\tilde{V}^{\Upsilon\eta_b}_{21}|_{S1}$ & $ -0.0032(3)$  & $0.0213$ & $0.1346$ & $-0.1634$ \\ 
    \hline \hline
  \end{tabular}
\end{center}
\end{table}

\begin{table}[t]
  \caption{ Values and correlation matrix elements of the $\tilde{V}^{\Upsilon\eta_b}_{21}|_i$, from set $4$ in Table \ref{tab:GluonEnsembles}. A value of $a^2q^2 = 0.00066(70)$ was found from the data.} 
\label{tab:ViSet4}
\begin{center}
  \begin{tabular}{ l | r | r r r }
    \hline \hline
    $p$ & Value & $C(p , \tilde{V}|_{F} )$ & $C(p , \tilde{V}|_{W1} )$  &  $C(p,\tilde{V}|_{S} )$ \\ \hline
    $\tilde{V}^{\Upsilon\eta_b}_{21}|_{F}$  & $0.1710(27)$     & & & \\ 
    $\tilde{V}^{\Upsilon\eta_b}_{21}|_{W1}$ & $ -0.0596(7)$    & $-0.4441$ &           &\\  
    $\tilde{V}^{\Upsilon\eta_b}_{21}|_{S}$  & $  -0.0289(10) $   & $-0.3281$ & $0.2708$  &\\
    $\tilde{V}^{\Upsilon\eta_b}_{21}|_{S1}$ & $ -0.0038(2)$      & $0.0206$ & $0.1493$ & $-0.3195$ \\ 
    \hline \hline
  \end{tabular}
\end{center}
\end{table}

\begin{table}[t]
  \caption{ Values and correlation matrix elements of the
    $\tilde{V}^{\Upsilon\eta_b}_{21}|_i$, from set $5$ in Table \ref{tab:GluonEnsembles}. A value of $a^2q^2 = -0.0021(6)$ was found from the data.} 
\label{tab:ViSet5}
\begin{center}
  \begin{tabular}{ l | r | r r r  }
    \hline \hline
    $p$ & Value & $C(p , \tilde{V}|_{F} )$ & $C(p , \tilde{V}|_{W1} )$  &  $C(p,\tilde{V}|_{S} )$ \\ \hline
    $\tilde{V}^{\Upsilon\eta_b}_{21}|_{F}$  & $0.1785(31)$     & & & \\ 
    $\tilde{V}^{\Upsilon\eta_b}_{21}|_{W1}$ & $-0.0618(15)$    & $-0.0703$ &           &\\  
    $\tilde{V}^{\Upsilon\eta_b}_{21}|_{S}$  & $-0.0276(10) $   & $-0.0925$ & $0.1526$  &\\
    $\tilde{V}^{\Upsilon\eta_b}_{21}|_{S1}$ & $-0.0060(5)$   & $0.0457$ & $0.3260$ & $-0.0266$ \\ 
    \hline \hline
  \end{tabular}
\end{center}
\end{table}

\begin{figure}[t]
  \centering
  \includegraphics[width=0.45\textwidth]{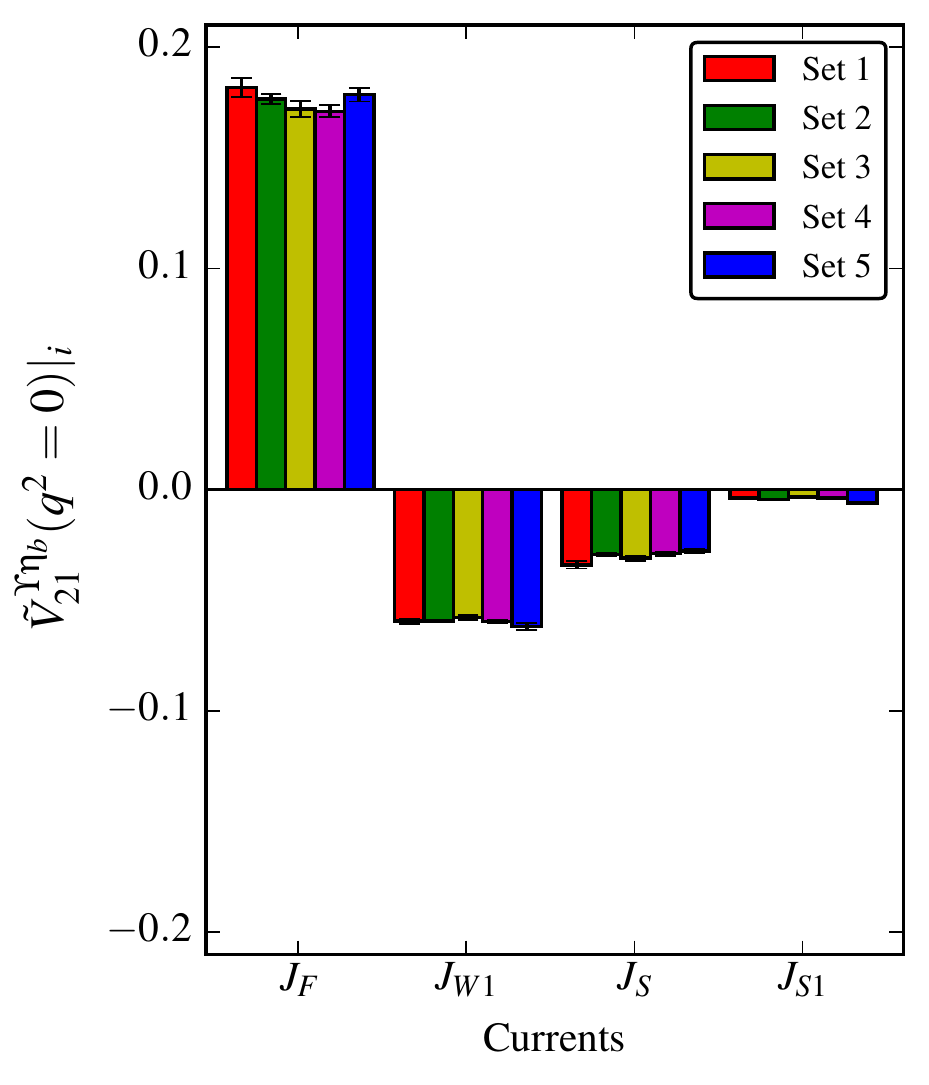}
  \caption{The value of the unrenormalised form factor, as described in the text,
    arising from each current across the different ensembles listed
    in Table \ref{tab:GluonEnsembles}. Statistical error only ( $\approx
    2-3\%$ for each current). }
  \label{fig:BarLatForm}
\end{figure}

The unrenormalised form factors, $\tilde{V}^{\Upsilon\eta}_{21}(q^2=0)|_i$, 
for each of the currents obtained from (\ref{eqn:CurrentsKN}) are computed for each of the ensembles listed in Table \ref{tab:GluonEnsembles} and their values are given in
Tables \ref{tab:ViSet1}, \ref{tab:ViSet2}, \ref{tab:ViSet3}, \ref{tab:ViSet4} and \ref{tab:ViSet5}. A
visual representation of $\tilde{V}^{\Upsilon\eta}_{21}(q^2=0)|_i$  is shown in
Figure \ref{fig:BarLatForm}. From this, we can see that the form factor from the current $J_F$ is leading order, and the other currents give a negative contribution
to $J_F$ of  approximately $ 30\%$, $20\%$, $3\%$ for $J_{W1}, J_{S},
J_{S1}$ respectively across all ensembles. Note that these values do not appear to obey the power counting for the currents given in Sec.~\ref{sec:Currents}; however, we understand (and explain below) that the leading-order contribution is suppressed for these hindered transitions. Similar
behaviour was seen in previous lattice NRQCD studies of this
decay \cite{Lewis:Rad1,Lewis:Rad2}. 

\begin{figure}[t]
  \centering
  \includegraphics[width=0.5\textwidth]{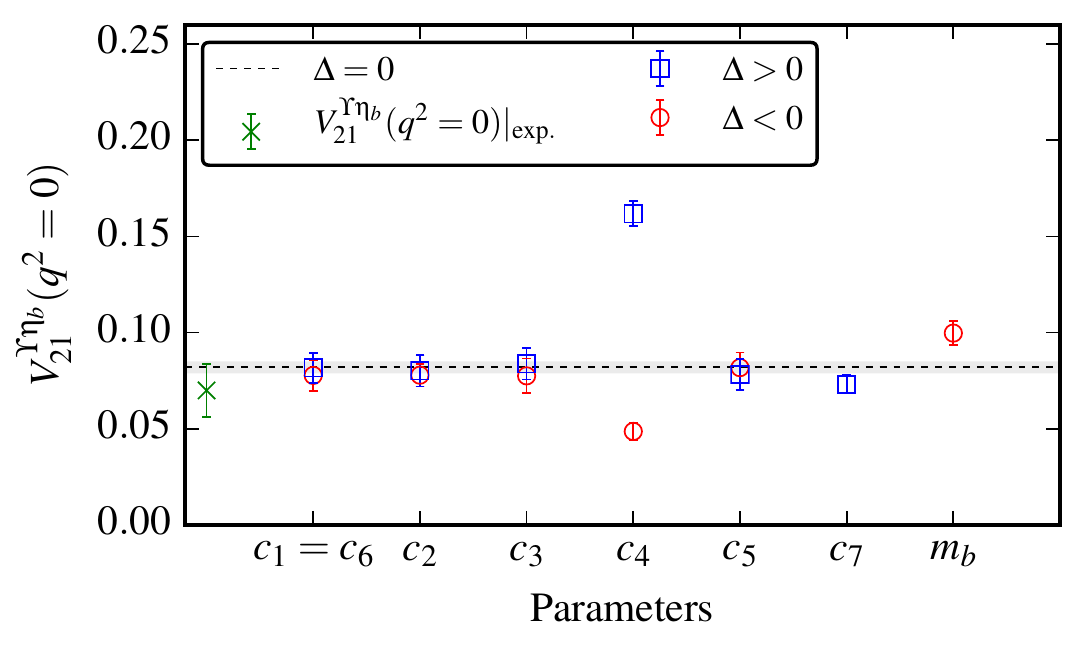}
  \caption{ The variation of the form factor with the parameters used in this study. $\Delta >0$ 
     ($\Delta<0$) denotes an upwards (downwards) shift in the parameter
    as described in the text, and the values of the varied parameters
    can be found in Table \ref{tab:VaryParams}. The data for $\Delta>0$ ($\Delta<0$) were generated on a subset of $400$ 
    configurations of the coarse lattice denoted Set 2 in Table
    \ref{tab:GluonEnsembles}. Statistical errors only. }
  \label{fig:Varyci}
\end{figure}

We also need to determine the
sensitivity of our form factors to the different 
parameters used in our calculation and use this analysis to give a reliable error budget. This is easily done in lattice NRQCD, as we
can simply change the value of a single parameter and rerun the
whole calculation. The results are shown in Figure
 \ref{fig:Varyci}, where we denote $p$ as a parameter to vary
 (either $c_i$ or $m_b$) and use $\Delta = p^{\text{test}} -
 p^{\mathcal{O}(\alpha_s)}$ to signify an upwards/downwards shift from
 the $\mathcal{O}(\alpha_s)$ correct value  $p^{\mathcal{O}(\alpha_s)}$ ($am_b$ is tuned fully nonperturbatively but we use $am_b = p^{\mathcal{O}(\alpha_s)}$ to avoid additional superfluous notation). The values of the changed parameters are given in Table \ref{tab:VaryParams}.

\begin{table}[t]
  \caption{ Values of the varied parameters used to obtain Figure
    \ref{fig:Varyci}. $\Delta >0$
    ($\Delta<0$) denotes an upwards (downwards) shift in the parameter
    as described in the text. $p^{\mathcal{O}(\alpha_s)}$ for $\Delta=0$ values are taken from Table \ref{tab:NRQCDParams} and reproduced here for convenience. }
\label{tab:VaryParams}
\begin{center}
  \begin{tabular}{|l | c | c| c | }
    \hline \hline
    Parameter & $p^{\text{test}}$ for $\Delta<0$ &
    $p^{\mathcal{O}(\alpha_s)}$ for $\Delta=0$ &   $p^{\text{test}}$ for $\Delta>0$     \\ \hline\hline
    $c_1=c_6$ & $1.00$ & $1.31$ & $1.50$ \\ \hline 
    $c_2$ & $0.75$ & $1.02$ & $1.25$  \\ \hline 
    $c_3$ & $0.75$ & $1.00$ & $1.25$  \\\hline 
    $c_4$ & $1.00$ & $1.19$ & $1.50$   \\\hline 
    $c_5$ & $1.00$ & $1.16$  & $1.50$ \\\hline 
    $c_7$ & $\overline{\hspace{0.7cm}}$ & $1.00$ & $1.50$  \\\hline 
    $m_b$ & $2.5935$  & $2.73$  & $\overline{\hspace{0.7cm}}$ \\
 \hline \hline
  \end{tabular}
\end{center}
\end{table}

\begin{figure}[t]
  \centering
  \includegraphics[width=0.45\textwidth]{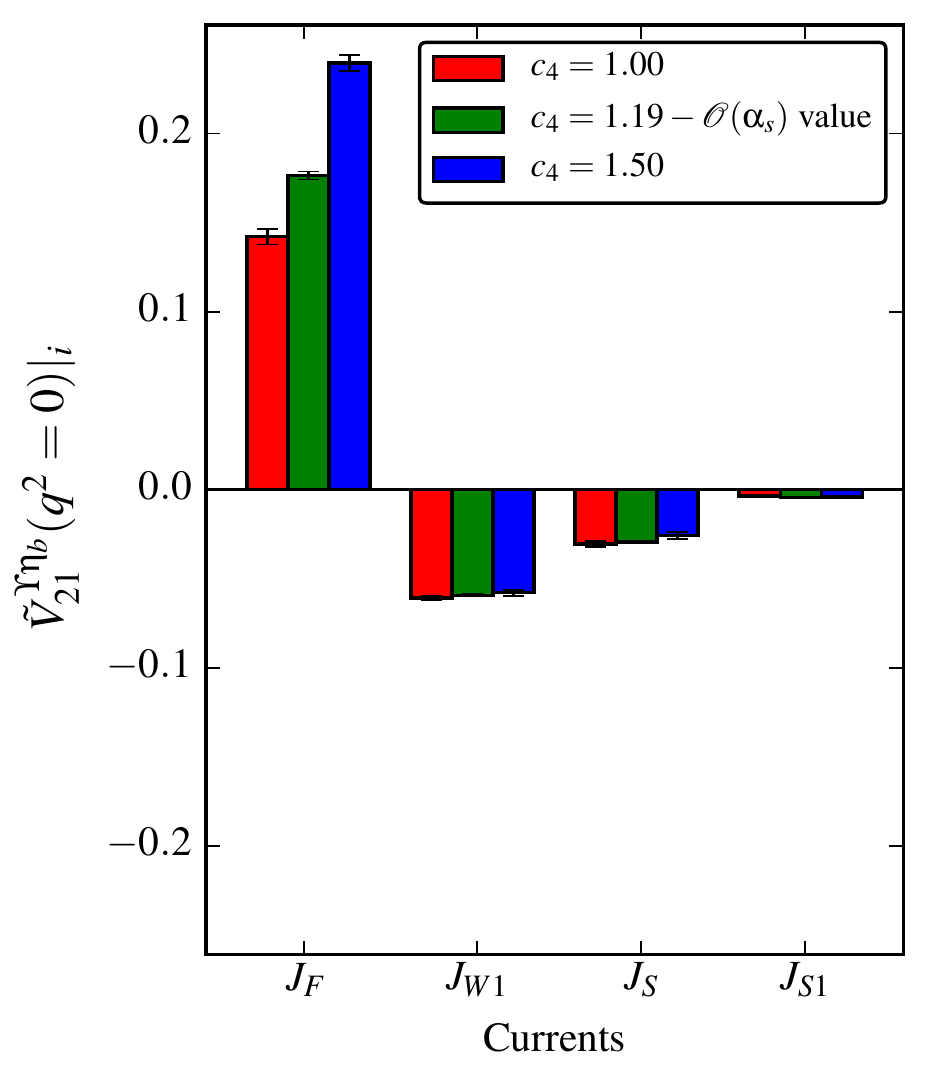}
  \caption{How each of the unrenormalised form factors from the different
    currents vary with $c_4$. As can be seen, the
    sensitivity comes from the $J_F$ current. The reason for this is
    described in Section \ref{sec:sensec4}. }
  \label{fig:Barc4Form}
\end{figure}

From Figure \ref{fig:Varyci} we can see that the form
factor is most sensitive to the value of $c_4$, while  $c_7$ and
$m_b$ are also important. We
need to describe this sensitivity in order to give a
reliable estimate on the error from not knowing each of these
parameters to infinite precision. Interestingly, it is useful to note that the sensitivity to
these parameters comes from the $J_F$ current, as shown in Figure
\ref{fig:Barc4Form}. We will use a simple potential model analysis to understand the deficiencies in the naive power counting, where these sensitivities arise from, and to gain insight into this hindered M$1$ decay. 

\subsection{Phenomenological Insight: Potential Model Analysis}
\label{sec:PotentialAnalysis}

In a potential model framework one would consider periodic harmonic time-dependent
perturbations and find the matrix element as the overlap between the
spatial part of the potential and the initial and final states under study. For an M$1$ decay,
mediated by either of the constituent quarks' magnetic moment $\VEC{\sigma}\cdot
\VEC{B}$, one can find the matrix element as \cite{QWG:2004} (labeling the spatial part of the potential as $J_F$, similar
to the current we use in Section \ref{sec:Currents} to highlight comparisons)
\begin{align}
&\langle  \eta_b(mS) | J_F | \Upsilon(nS) \rangle  =  \nonumber \\ 
& \hspace{1.5cm} \mathcal{S}_{fi} \int_0^{\infty} dr ~ r^2 R^*_{m,\eta_b}(r) j_0\left(\frac{|q|r}{2}\right) R_{n,\Upsilon}(r)  \nonumber \\ 
\intertext{with the integral expanded as}
&\int_0^{\infty} dr ~ r^2 R^*_{m,\eta_b}(r)
j_0\left(\frac{|q|r}{2}\right) R_{n,\Upsilon}(r)   = \nonumber \\ 
& \hspace{1.5cm} \delta_{nm} + a_2|q_{\gamma}|^2 r_0^2 + a_4|q_{\gamma}|^4 r_0^4 + \cdots \,.
\label{eqn:WavefunctionOverlap}
\end{align}
Here, we have factored the spin piece $\mathcal{S}_{fi}$ in the matrix element from 
the radial integral (appropriate in the nonrelativistic limit) and used the Taylor
expansion of $ j_0(x) = {\sin(x)}/{x} = \sum_n {(-1)^n
  x^{2n}}/{(2n+1)!}$ to see that it is a polynomial in
$|q_{\gamma}|^2$. Additionally, the only scale in the wavefunctions
capable of being combined with $|q_{\gamma}|^2$ to make it
dimensionless  is some combination of the Bohr radii of each state,
which we call $r_0$. The $a_{2l}$ are coefficients which could be
calculated if wave-functions were supplied. The leading Kronecker
$\delta$-function in (\ref{eqn:WavefunctionOverlap}) comes from noting
the orthogonality condition in the extreme nonrelativistic limit,
$|q_{\gamma}|^2 \to 0$. 

As can be seen, for a $nS \to nS$ transition, the leading order term in
(\ref{eqn:WavefunctionOverlap}) is one. However, for transitions between different radial excitations, the $\delta_{nm}$ vanishes and we are left with
a leading order $|q_{\gamma}|^2 r_0^2$ term. The radii of the
bottomonium states under study are of the
order the reciprocal of the typical momentum, e.g, $r_0 \sim
1/mv$. Thus, as $|q_{\gamma}|^2 r_0^2 \sim m^2 v^4 /(m^2 v^2) \sim
v^2$, the leading order matrix element from $J_F$ in a radially excited decay is
suppressed by a factor of $v^2$ more than naively expected from using
power-counting rules on the currents alone. This suppression leads to
an array of sensitivities that make this decay particularly difficult
to pin down theoretically from within a potential model
\cite{Godfrey:2001}, as we will expand upon in Section \ref{sec:Conclusions}. 

Due to the derivatives in the other currents listed in
(\ref{eqn:CurrentsKN}), the matrix elements of these currents give rise
to wavefunction overlaps that are not orthogonal in the extreme
nonrelativistic limit, and as such are not more suppressed for radially
excited transitions. The derivatives act on the initial bottomonium
state and give a leading order $p \sim \mathcal{O}(mv)$ effect, which
does not depend on the photon momentum, as can be seen by taking the
$|q_{\gamma}| \to 0$ limit.  This results in the relativistic
corrections to the leading order $J_F$ current, which we have included
in our calculation, having appreciable 
effects (see Fig.~\ref{fig:BarLatForm}), namely $J_{W1}$, $J_S$.  The orthogonality of the radial wavefunction muddles up the power counting of the first few currents, but additional derivatives in relativistic corrections to these currents will suppress them further. By including the current $J_{S2}$, we check that added derivitives do suppress the contribution of the current further
as expected. 

By examining (\ref{eqn:WavefunctionOverlap}), we found that the leading order matrix element for
the radially-excited radiative  transition can be suppressed more 
than we would naively expect from just power-counting the current alone.
Relativistic corrections to the $J_F$ current are then appreciable,
explaining the behaviour seen in Figure \ref{fig:BarLatForm}.  Even if we included the relativistic
corrections to the current in a potential model, we still would not get the
correct value for this decay, as we also need to consider all relativistic
corrections to the wavefunctions arising from perturbative potentials in
the Hamiltonian. This gives rise to the sensitivities to the different parameters as seen
in Figure \ref{fig:Barc4Form}, which we explain below. To do so,  it is sufficient to consider first order time-independent perturbation theory.

\subsection{Sensitivity and Errors from Terms in the NRQCD Action}
\label{sec:PertAnalysis}

We want to consider potentials arising from relativistic corrections in the NRQCD action causing perturbations of the wavefunction. To first order in $\alpha_s$ we have
\begin{align}
& | \eta_b( 1S) \rangle ^{(1)} = | \eta_b( 1S) \rangle ^{(0)} - \sum_{m \ne
1} | \eta_b( mS) \rangle ^{(0)} \frac{ V_{m1}^{\eta_b} }{E^{\eta_b}_{m1}}
\nonumber \\
& | \Upsilon( 2S) \rangle ^{(1)} = | \Upsilon( 2S) \rangle ^{(0)} - \sum_{n \ne
2} | \Upsilon( nS) \rangle ^{(0)} \frac{ V_{n2}^{\Upsilon}
}{E^{\Upsilon}_{n2}} \,.
\label{eqn:PertStates}
\end{align}
The state $| n \rangle^{(1)}$ ( $| n
\rangle^{(0)}$) is the first-order perturbed state (the unperturbed state), $V_{nm} =$ $ ^{(0)}\langle n |
V | m \rangle ^{(0)}$ with $V$ being the potential representing the perturbation and $E_{nm} = E_n^{(0)} - E_m^{(0)}$. Now, we take currents of interest between these states to yield
\begin{align}
&^{(1)}\langle  \eta_b(1S) | J_i | \Upsilon(2S) \rangle ^{(1)}  =
\nonumber \\ 
&\hspace{1.5cm}  ^{(0)} \langle \eta_b(1S) | J_i | \Upsilon(2S) \rangle^{(0)} \nonumber \\ 
&\hspace{1.5cm} - \sum_{m
  \ne 1 } \frac{{V_{m1}^{\eta_b}}^*}{E^{\eta_b}_{m1}} ~^{(0)}\langle \eta_b(mS) | J_i | \Upsilon(2S) \rangle^{(0)} \nonumber \\
&\hspace{1.5cm} - \sum_{n
  \ne 2 } \frac{ {V_{n2}^{\Upsilon}} }{E^{\Upsilon}_{n2}} ~^{(0)}\langle
\eta_b(1S) | J_i | \Upsilon(nS) \rangle^{(0)}  \,.
\label{eqn:InteractingMatrix}
\end{align}

As mentioned above, for the current $J_F$,  due to the fact that $^{(0)} \langle \eta_b(1S) | J_F | \Upsilon(2S) \rangle^{(0)}$ is suppressed for radially excited decays, the $^{(0)} \langle \eta_b(nS) | J_F | \Upsilon(nS) \rangle^{(0)}$ pieces in the second term in (\ref{eqn:InteractingMatrix}) become appreciable. The matrix elements arising from currents with derivatives are already
suppressed, and  the first order corrections to these matrix elements 
are not appreciable, as seen in Figure \ref{fig:Barc4Form}.

\subsubsection{Sensitivity and Error from $c_4$:}
\label{sec:sensec4}

Including a potential from
the exchange of a single gluon between two vertices involving the
chromomagnetic operator as shown in Appendix \ref{app:Potential}, we find
\begin{align}
& \hspace{-1cm} ^{(1)}\langle \eta_b(1S) | J_F | \Upsilon(2S) \rangle^{(1)}  =
\nonumber \\
&  ^{(0)}\langle \eta_b(1S) | J_F | \Upsilon(2S)
\rangle^{(0)} +  \frac{c_4^2 g^2 }{9m_b^2 E_{21}} \psi^*_1 (0) \psi_2(0) \nonumber
\\
& \times \big( 6  ~^{(0)}\langle
\eta_b(2S) | J_F | \Upsilon(2S) \rangle ^{(0)} \nonumber \\ 
& \hspace{1cm}+ 2 ~^{(0)}\langle
\eta_b(1S) | J_F | \Upsilon(1S) \rangle ^{(0)}  + \mathcal{O}(v^2)
\big) \nonumber \\ 
= &  ~  ^{(0)}\langle  \eta_b(1S) | J_F | \Upsilon(2S) \rangle ^{(0)} \nonumber \\ 
& + \frac{8 c_4^2 g^2 }{9m_b^2 E_{21}} \mathcal{S}_{fi}\psi^*_1 (0) \psi_2(0) +
\mathcal{O}(v^2) \,. \label{eqn:c4reduced}
\end{align}

\begin{table*}
  \caption{ Values of the form factor $\tilde{V}^{\Upsilon\eta_b}_{21}|_F$ with a
    variation of certain parameters from the lattice NRQCD data on a coarse
    lattice (Set $2$ in Table \ref{tab:GluonEnsembles}). Error is statistical only. }
\label{tab:VFVaryCoarse}
\begin{center}
  \begin{tabular}{ c | c | c c c c  }
    \hline \hline
    $p$ & Value & $C(p , \tilde{V}^{\Upsilon\eta_b}_{21}|_{F,c_4=1.00} )$ & $C(p ,
    \tilde{V}^{\Upsilon\eta_b}_{21}|_{F,c_4=1.19} )$  &
    $C(p,\tilde{V}^{\Upsilon\eta_b}_{21}|_{F,c_7=1.50} )$ &  $C(p,\tilde{V}^{\Upsilon\eta_b}_{21}|_{F,c_2=1.25} )$ \\ \hline
    $\tilde{V}^{\Upsilon\eta_b}_{21}|_{F,c_4=1.00}$ & $0.1426(47)$  & & && \\ 
    $\tilde{V}^{\Upsilon\eta_b}_{21}|_{F,c_4=1.19}$ & $0.1772(44)$  & $0.3040$ &   &&\\  
    $\tilde{V}^{\Upsilon\eta_b}_{21}|_{F,c_7=1.50}$ & $0.1687(67) $ & $0.0342 $ & $0.0472$  &&\\
    $\tilde{V}^{\Upsilon\eta_b}_{21}|_{F,c_2=1.25}$ & $0.1769(46)$ & $0.2979$ &
    $   0.3352 $ & $ 0.0467 $& \\ 
    $\tilde{V}^{\Upsilon\eta_b}_{21}|_{F,m_b=2.59}$ & $0.1939(48)$ & $0.3070$
    & $0.3479 $  & $ 0.0508 $ & $ 0.3411$\\
    \hline \hline
  \end{tabular}
\end{center}
\end{table*}

The reason for the sensitivity to $c_4$ is clear. The matrix element $~^{(0)}\langle
\eta_b(1S) | J_F | \Upsilon(2S) \rangle ^{(0)}$ is suppressed due to the
orthogonality of the radial wavefunctions in
(\ref{eqn:WavefunctionOverlap}), while $~^{(0)}\langle
\eta_b(nS) | J_F | \Upsilon(nS) \rangle ^{(0)}$ is not. This results in
the second term in (\ref{eqn:c4reduced}) being
sizeable compared to the first.

Since we have values of the form factor at three values of $c_4$ on a
coarse lattice as shown in Figure \ref{fig:Barc4Form}, and an
understanding that the functional dependence of the form factor on $c_4$ should be $\tilde{V}^{\Upsilon}_{F}  = a_{c_4} + c_4^2
b_{c_4}$ from (\ref{eqn:c4reduced}), we should check that this is
consistent. We use the $c_4=1.00$ and $c_4=1.19$ values from our lattice
NRQCD calculation listed in Table \ref{tab:VFVaryCoarse} to find the values of
$a_{c_4}$ and $b_{c_4}$ in Table \ref{tab:VFDependCoarse}. 

\begin{figure}[t]
  \centering
  \includegraphics[width=0.5\textwidth]{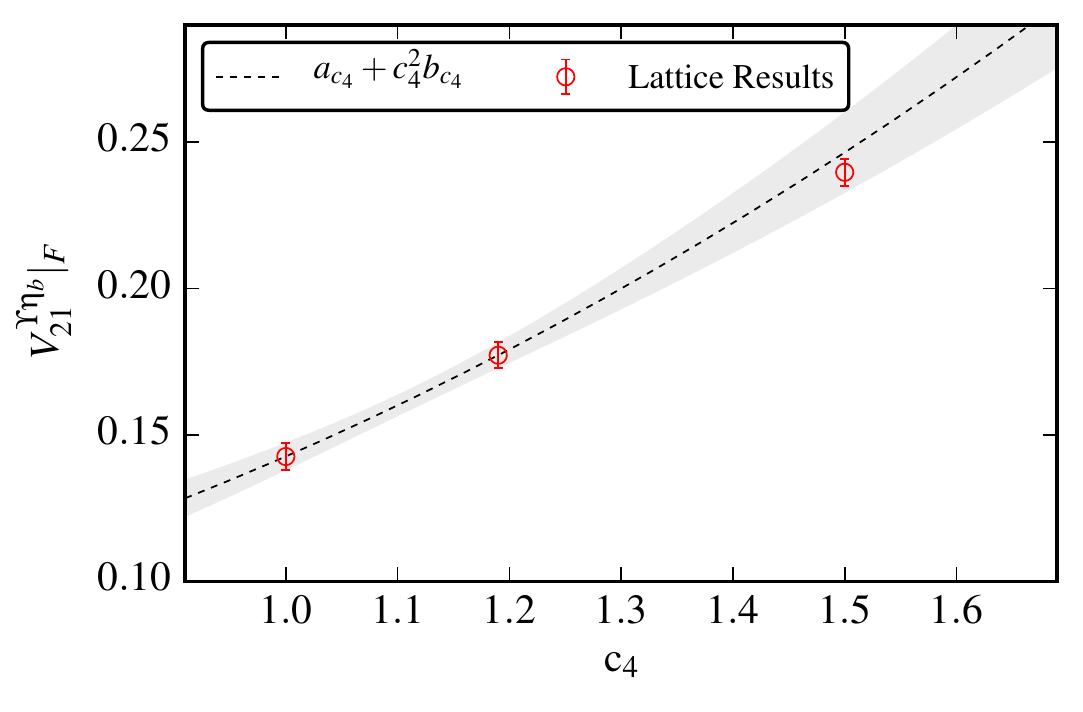}
  \caption{ The $c_4$ dependence of $\tilde{V}^{\Upsilon\eta_b}_{21}|_{F}$ as described in the text, along with the lattice values of $\tilde{V}^{\Upsilon\eta_b}_{21}|_{F}$. }
  \label{fig:VaryC4}
\end{figure}

We can also relate the second term from the leading order approximation in (\ref{eqn:c4reduced}) to quantities that are measured in experiment and check the consistency of the value of $b_{c_4}$ given in Table \ref{tab:VFDependCoarse}. By comparing the decay rate formulae from a potential model calculation \cite{Dudek:CharmRad} with the one given in (\ref{eqn:UpsilonDecayRate}), we find:
\begin{align*}
V^{\Upsilon\eta_b}_{21} & = \left( \frac{m_{\Upsilon(2S)} + m_{\eta_b(1S)}}{2m_b} \right) \\ 
& \hspace{1cm} \times \int_0^{\infty} dr ~ r^2 R^*_{m,\eta_b}(r) j_0\left(\frac{|q|r}{2}\right) R_{n,\Upsilon}(r)
\end{align*}
and then using this in (\ref{eqn:c4reduced}) yields: 
\begin{align}
c_4^2b_{c_4} & = \left( \frac{m_{\Upsilon(2S)} + m_{\eta_b(1S)}}{2m_b}\right) \frac{\sqrt{\Delta(2S)\Delta(1S)}}{E_{21}} + \mathcal{O}(v^2)
\label{eqn:bc4Exp}
\end{align}
where $\Delta(iS)$ is the hyperfine splitting between $i$'th radial excitations. Using the values of $c_4$, $a$ and $am_b$ from set $2$ in Table \ref{tab:NRQCDParams}, along with the PDG average \cite{PDG:2014} values for $\Delta(iS)$ and the spin averaged $E_{21}$, we find $b_{c_4} = 0.105(14)$. This is consistent with the value of $b_{c_4}$ from Table \ref{tab:VFDependCoarse}. 

In Figure \ref{fig:VaryC4}, we show the strong $c_4$ dependence of $\tilde{V}^{\Upsilon\eta_b}_{21}|_{F}  = a_{c_4} + c_4^2 b_{c_4}$, along with the the lattice values of $\tilde{V}^{\Upsilon\eta_b}_{21}|_{F}$ shown in Figure \ref{fig:Barc4Form}. This illustrates both the need for at least the $\mathcal{O}(\alpha_s)$-correct value of $c_4$ and the consistency of $a_{c_4}$ and $b_{c_4}$ with all our lattice data.

Since we only know $c_4$ to one loop in perturbation theory, there will
be a systematic error associated with not knowing it to higher
orders. With the above functional dependence of $\tilde{V}^{\Upsilon\eta_b}_{21}|_{F}  = a_{c_4} + c_4^2 b_{c_4}$, an error of $2\alpha_s^2b_{c_4}$ should be introduced from
not knowing $c_4$ to second order. As there is little lattice spacing dependence in the unrenormalised form factors as shown in Figure \ref{fig:BarLatForm}, we use the value of $b_{c_4}$ from Table \ref{tab:VFVaryCoarse} across all ensembles and  introduce an additive systematic error (correlated across
lattice spacings) of $2\alpha_s^2 b_{c_4}$  from not knowing
$c_4$ to more than one loop. We also allow for the statistical error
in the determination of $c_4^{(1)}$ coming from the Vegas integration
\cite{Hammant:2013} by adding an error of $2\alpha_s \delta c_4^{(1)} b_{c_4}$. 

With the other currents that have derivatives, the situation is significantly different. Due to the
derivatives, the second term in (\ref{eqn:InteractingMatrix}) is always suppressed
and relativistic corrections are not an appreciable
effect, as seen in Figure \ref{fig:Barc4Form}. Variations of these currents with $c_4$ are not appreciable within the other errors.

\subsubsection{Sensitivity and Error from $c_7$: }
\label{sec:sensec7}

The $c_7$ operator is a $D^2$ correction to the $c_4$ term and is expected to be a
$\mathcal{O}(v^2)$ effect. We can proceed as before,
assuming a linear functional dependence on $c_7$ as  $\tilde{V}^{\Upsilon\eta_b}_{21}|_F = a_{c_7}
+ c_7 b_{c_7} $, coming from the exchange of a single gluon from a $c_4$
vertex and a $c_7$ vertex. Using our data points in Table
\ref{tab:VFVaryCoarse}, we find $a_{c_7}, b_{c_7}$ listed in Table \ref{tab:VFDependCoarse}. 

It is seen that $b_{c_7}$ gives a negative contribution as a consequence of the
$D^2$ and the ratio $b_{c_7}/b_{c_4} = -0.20(18)$ should be a
$\mathcal{O}(v^2)$ effect. This is roughly consistent. We
assume a dependence on $c_7$ as $b_{c_7} \approx 2v^2b_{c_4} = 0.2b_{c_4}$, and similarly
to the $c_4$  error  above, introduce an
additive systematic error (correlated across lattice spacings) of
$2\alpha_s v^2 b_{c_4}$ from not knowing $c_7$ past tree-level.  Just as with variations of $c_4$, the currents with  derivatives are
insensitive to variations of $c_7$ and are all consistent within our
small statistical errors. 

\subsubsection{Sensitivity and Error from $m_b$:}
\label{sec:sensemb}

Using the fact that radial splittings are expected to be $E_{21} \sim
m_bv^2$, by examining (\ref{eqn:c4reduced}) we observe that the form factor should
have a functional dependence on $m_b$ as $\tilde{V}^{\Upsilon\eta_b}_{21}|_F = a_{m_b} + b_{m_b}/m_b^3$.   Using our data points in Table
\ref{tab:VFVaryCoarse}, we find $a_{m_b}, b_{m_b}$ listed in Table
\ref{tab:VFDependCoarse}. 

Again, we can check consistency within this first order
approximation. Comparing the assumed functional forms against the
equation from which they came (\ref{eqn:c4reduced}), we find $ b_{m_b}
= c_4^2 m_b^3 b_{c_4}$. Thus, using the values of $b_{c_4}, b_{m_b}$
we obtain from the lattice data, we find the ratio $b_{m_b}
/ c_4^2 m_b^3 b_{c_4} =  0.85(35)$, consistent with $1.0$. 

We allow for a systematic error from the (small) uncertainty in
mistuning the $b$-quark
mass estimated from \cite{Dowdall:Upsilon}. By using the
above inverse cubic functional dependence on $m_b$, we find of an error of
$3b_{m_b} \delta_{m_b}/ m_b^4$. Using the estimate of
$b_{m_b}$ in terms of $b_{c_4}$, we find the error as $3 c_4^2  b_{c_4} \delta_{m_b}/m_b$.

\begin{table*}
  \caption{ Values of the functional dependency of
    $\tilde{V}^{\Upsilon\eta_b}_{21}|_F$ with parameters from the action using
    data from Table \ref{tab:VFVaryCoarse}. See text for details. Error is statistical only. }
\label{tab:VFDependCoarse}
\begin{center}
  \begin{tabular}{ c | r | r r r r r r r }
    \hline \hline
    $p$ & Value & $C(p , a_{c4} )$ & $C(p , b_{c4} )$  & $C(p , a_{c7}
    )$ &  $C(p , b_{c7} )$ & $C(p , a_{c2} )$  & $C(p , b_{c2} )$ &
    $C(p , a_{m_b} )$  \\ \hline
    $a_{c4}$ & $0.060(16)$  & & && &&\\ 
    $b_{c4}$ & $0.083(13)$  & $-0.974$ &   &&&&\\  
    $a_{c7}$ & $0.194(18)$ & $-0.257 $ & $  0.395$  &&&&\\
    $b_{c7}$ & $-0.017(16)$ &  $0.202 $ & $-0.307$& $-0.979$&& \\ 
    $a_{c2}$ & $0.179(24)$ & $-0.389$ & $  0.510 $  & $  0.486$ & $  -0.378$&&\\
    $b_{c2}$ & $-0.001(21)$ & $0.366$ & $-0.459$  & $ -0.403 $ & $ 0.316$& $ -0.988$ &\\
    $a_{m_b}$ & $0.077(34)$ & $-0.382$ & $   0.487$  & $0.442 $ & $-0.346$& $ 0.562$ & $ -0.506$ \\
    $b_{m_b}$ & $2.04(65)$ & $ 0.363$ & $-0.448$  & $ -0.381$ & $0.300$& $-0.512$ & $ 0.469$ & $ -0.994$\\
    \hline \hline
  \end{tabular}
\end{center}
\end{table*}

\subsubsection{Sensitivity and Error from $c_2$:}
\label{sec:sensec2}
From our numerical data, it appears as if the form factor is not
sensitive to a variation in $c_2$. We can understand this and use it in
our analysis of the errors. In Appendix \ref{app:Potential} we show how the the leading spin-independent perturbative potential from the exchange of a single gluon involving the Darwin
term at one of the vertices \cite{Hammant:2013} gives rise to a correction to the leading order matrix element that is  $\mathcal{O}(\alpha_s v^2)$. Using the data in Table
 \ref{tab:VFVaryCoarse} for how $\tilde{V}^{\Upsilon\eta_b}_{21}|_F$ varies with
 $c_2$, and using the functional form $\tilde{V}^{\Upsilon\eta_b}_{21}|_F = a_{c_2} +
 c_2 b_{c_2}$, we find the values listed in Table \ref{tab:VFDependCoarse}.  

To test the consistency of this description, by comparing the value
$b_{c_4}$ associated with the second term in (\ref{eqn:c4reduced}) and
the second term in (\ref{eqn:c2reduced}) we see $b_{c_2} \approx
3v^2 b_{c_4} / 8$. Using the values in Table \ref{tab:VFDependCoarse} gives $3v^2 b_{c_4} / 8 = 0.00311(49)$, consistent with $b_{c_2} = 0.001(21)$. Due
to the smallness of this dependency, we can safely neglect the
systematic error from not knowing $c_2$ to two loop order.

\subsubsection{Sensitivity and Error from $c_3$:}

Since the bottomonium states under study have no orbital angular
momentum, there is no sensitivity to $c_3$ arising from a spin-orbit
perturbing potential. This is confirmed by the numerical data in
Figure \ref{fig:Varyci}. We introduce no error from $c_3$.

\subsubsection{Sensitivity and Error from Four-Quark Operators:}
\label{sec:sense4q}

The four quark operators in NRQCD \cite{Dowdall:Upsilon} are contact terms between the quark and
anti-quark fields arising from $\alpha_s^2$ processes in relativistic
QCD. These can have a noticable effect on the
hyperfine splitting \cite{Dowdall:Hyperfine}. Since the matrix element in
(\ref{eqn:c4reduced}) is sensitive to parameters in much the same way as the
hyperfine splitting, we would expect contributions from the four quark
operators. In Appendix \ref{app:Potential}, we show the effect of the four-quark potential on the matrix element to first order. 

We introduce a systematic error (correlated across lattice sites) for
neglecting these leading order four quark operators in our
calculation. We estimate this by comparing the second term in
(\ref{eqn:c4reduced}) with the second term in (\ref{eqn:4qreduced})
 to find an
error $27 b_{c_4}(d_1\alpha_s - d_2\alpha_s)/16\pi$ and then use the values  of $d_1\alpha_s - d_2\alpha_s$ from \cite{Hammant:2013} (as corrected per \cite{Ron:FQ}).

\subsubsection{Error from Missing Higher Order Operators in the NRQCD Action:}
\label{sec:errorv8}

The terms in the action that have not been considered are the
$\mathcal{O}(v^2)$ corrections to the $c_2$ and $c_7$ terms. Since the
coefficient $b_{c_2}$ is already quite small, the
$v^2$ correction to this will be negligible within our numerical
precision and can be neglected. The error from $v^2$ corrections to
$c_7$  is estimated as $v^2b_{c_7} = 2v^4 b_{c_4}$.

\subsubsection{Total Error on $\tilde{V}^{\Upsilon\eta_b}_{21}|_F$ from Terms in the NRQCD Action:}
\label{sec:ErrVF}

After performing the final continuum and chiral extrapolation as shown in Section \ref{sec:fullerror}, we can obtain a breakdown of how each of the uncertainties arising from the NRQCD action effects the error in $\tilde{V}^{\Upsilon\eta_b}_{21}|_F$ as a percentage of the error on the total form factor given in Table \ref{tab:ErrorBudgetUps}. We find that the errors from the NRQCD action contribute to a $10.4\%$ systematic error in $\tilde{V}^{\Upsilon\eta_b}_{21}|_F$ as a percentage of the total error on the total form factor. In order of dominance, the most sizable of these errors is a  $7.9\%$ error from neglecting the $\mathcal{O}(\alpha_s^2)$ correction in $c_4$, then a $4.4\%$ error from the statistical error in $c_4^{(1)}$ while $3.9\%$ comes from neglecting the one-loop correction to $c_7$. These numbers should be added in quadrature and each is a percentage of the total error on the total form factor.

 Note that due to the destructive interference between the leading order form factor, $\tilde{V}^{\Upsilon\eta_b}_{21}|_F$, and the other currents as shown in Section \ref{sec:DecayNRQCD}, the error coming from $\tilde{V}^{\Upsilon\eta_b}_{21}|_F$ as a percentage of the total error on ${V}^{\Upsilon\eta_b}_{21}$ is larger than the errors on $\tilde{V}^{\Upsilon\eta_b}_{21}|_F$ alone. As a result, improvement of the errors coming from the NRQCD action has an appreciable effect.

\subsubsection{Test of Uncertainties from the NRQCD Action:}
\label{sec:HyperTest}

To ensure that we have performed a reasonable estimation of the
errors arising from the NRQCD action, we have also tuned $c_4$ against the $\Upsilon(1S) - \eta_b(1S)$
hyperfine splitting on the coarse lattice denoted set $2$ in Table
\ref{tab:GluonEnsembles}. In a perturbative framework as
described above, the hyperfine splitting can be pictured as 
a result of perturbative potentials shifting the unperturbed
energies. The most sizable of these is the leading order $c_4^2$
potential, as described in Section \ref{sec:sensec4}, and then the
four-quark potential, as described in Section \ref{sec:sense4q}.  In a numerical
calculation with no four-fermion operators,  tuning
the numerical  hyperfine splitting against the experimental one
would have the effect of absorbing the above four-fermion term (among
others) into the tuned $c_4$. Stated more concretely, 
\begin{align}
\left(c_4^{\text{lat}}\right)^2 \to \left(c_4^{tuned}\right)^2 & = c_4^2 - \frac{27}{16\pi} (d_1 -
d_2)\alpha_s \,. \label{eqn:c4tuned}
\end{align}

Then, putting $\left(c_4^{tuned}\right)^2$ into (\ref{eqn:c4reduced}) gives
exactly the four fermion term which we need in (\ref{eqn:4qreduced}). As such, using
$c_4^{tuned}$ numerically would include the effect of
the four fermion operator for this decay automatically. For the nonperturbative tuned
$c_4^{tuned}$ error budget, there are no $c_7$, leading order
four-quark,  or missing $v^8$ operator errors as these will be absorded into
the value of $c_4^{tuned}$ and fed back into the matrix element calculation
automatically.  However,
from (\ref{eqn:4qreduced}) we see there is still an additive
systematic error of  $3v^2(27/16\pi)\alpha_s b_{c_4}$ from only
knowing the difference $(d_1 - d_2)$, and not $d_1$ and
$d_2$ individually. 

The Particle Data Group average for the hyperfine
splitting is $\Delta^{\text{exp.}} = 62.3(3.2)$ MeV \cite{PDG:2014}, while our lattice
calculation with $c_4 = 1.23$ gives $\Delta^{\text{lat}} = 62.54(46)$ MeV (statistical and
scale setting error only). We
get a value of $c_4^{\text{tune}} = 1.230(5)(31)$ from tuning $c_4$
against the experimental hyperfine splitting, where the first error is
from the lattice, and the second from experiment.  The change from the one-loop perturbative value $1.19$ to the nonperturbatively tuned $1.230(5)(31)$ is well-accounted for in the error budget (see Sec.~\ref{sec:ErrVF}) from the statistical error on
$\delta c_4^{(1)}$ alone, and including the higher order corrections
to $c_4$ and the four-quark error is significantly over-compensating for this
change. 

Rerunning the computation of the form factor with $c_4 = 1.23$, gives
a value of $V^{\Upsilon\eta_b}_{21} = 0.097(14)$. This includes all errors,
and the only difference from the above error budget is that the error in
$\tilde{V}^{\Upsilon\eta_b}_{21}|_F$ now comes from $c_4^{\text{tune}}$ and the error from
knowing only the difference $d_2 -d_1$. This value is to be compared
with the form factor from a perturbatively tuned $c_4$  shown in Section
\ref{sec:fullerror}, i.e., $V^{\Upsilon\eta_b}_{21} = 0.089(22) $. These are entirely consistent, giving evidence
that our error budget is a reliable estimation of the errors.

The four-quark operators appear to increase the value of
the form factor, in a similar way as they do for the hyperfine
splitting. However, it was found that
including the four-quark operators in the calculation of the hyperfine
splitting largely changed the slope of the
continuum extrapolation but did not shift the final result
away from the value computed without the four-fermion operators included
\cite{Dowdall:Hyperfine}.

Based on our analysis, we estimate that by tuning $c_4$ against the hyperfine splitting on all ensembles and re-doing the full calculation, one could reduce the error on $\tilde{V}^{\Upsilon\eta_b}_{21}|_F$ to $\sim 4\%$. Also, we estimate that such a calculation would give an error on the final form factor of $\sim 11\%$ (compared against the value given in Table \ref{tab:ErrorBudgetUps}), where now the uncertainties in order of dominance are from the neglected currents, neglecting the mixing down in $\omega^{(1)}_F$, and neglecting the one-loop correction to $\omega_{W1}$.

\subsection{Errors from Missing Higher Order Currents}
\label{sec:errorcurrents}

Since we are using an effective field theory to study this transition, there will be
higher order currents which we have
not included in this study  but that contribute to the final form factor. The most sizable current which we have not 
included is the $D^2$ correction to $J_{W1}$.  Therefore, we include a systematic
uncertainty (correlated across all lattice sites) of $v^2\tilde{V}^{\Upsilon\eta_b}_{21}|_{W1}$.  

\subsection{Full Error Budget}
\label{sec:fullerror}

\begin{figure}[t]
  \centering
  \includegraphics[width=0.495\textwidth]{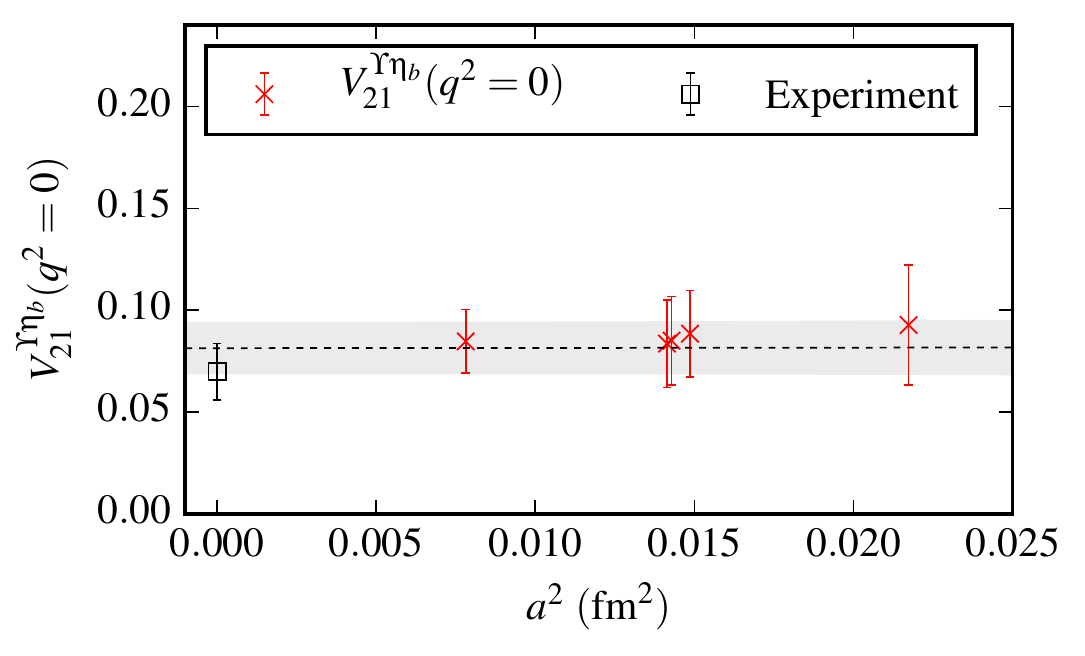}
  \caption{ Fit results for the form factor relevant to the $\Upsilon(2S)\to\eta_b(1S)\gamma$
    decay. All errors included. The error budget is shown in Table
    \ref{tab:ErrorBudgetUps}. }
  \label{fig:FullConExtrapUps}
\end{figure}

After the analysis performed in the previous sections, we are now in a
position to give a full error budget for the form factor
$V^{\Upsilon\eta_b}_{21}$. To compare to experiment, we perform a simultaneous
lattice spacing and sea quark mass extrapolation. We fit results from
all ensembles to the form \cite{Dowdall:Upsilon, Colquhoun:2015}
\begin{align}
V(a^2, am_b)  = & V_{\text{phys}} \times \Big [ 1 \nonumber \\
&+ \sum_{j=1,2}(a\Lambda)^{2j} k_j\left(1 + k_{j1}\delta x_m +
  k_{j2}(\delta x_m)^2 \right) \nonumber \\
&+ 2l_{1} \delta m \left(1 + l_2 (a\Lambda)^2 \right) \Big ] \,.
\label{eqn:FitFunction}
\end{align}
The lattice spacing dependence is set by a scale $\Lambda = 500$ MeV,
$\delta x_m = (am_b -2.7)/1.5$ allows for a mild dependence on the
effective theory cutoff $am_b$, and $\delta x_l = (am_l/am_s) -
(am_l/am_s)_{\text{phys}}$ for each ensemble with
$(m_l/m_s)_{\text{phys}} = 27.4(1)$ is taken from lattice QCD
\cite{mlms:2014}. We take a Gaussian prior on the leading order $a^2$ term to be $0.0(3)$, as
the HISQ action is correct through $\mathcal{O}(\alpha_sa^2)$; a prior
of $0.0(1.0)$ on the higher order $a$ terms; a prior of $0.00(3)$ on
$l_1$ allowing for a $3\%$ shift if the light quarks were as heavy as
the strange; a prior of $0.10(5)$ on $V_{\text{phys}}$\footnote{The width on this prior is chosen so as to ensure that the fitted result is insensitive to the central value.}. The extrapolation with all errors is shown in Figure \ref{fig:FullConExtrapUps} and a
full error budget is shown in Table \ref{tab:ErrorBudgetUps}. 

\begin{table}[t]
  \caption{Full error budget for the total form factor ${V}^{\Upsilon\eta_b}_{21}$  relevant for the $\Upsilon(2S)\to\eta_b(1S)\gamma$ decay from Figure \ref{fig:FullConExtrapUps}.  A discussion of the uncertainties in $\tilde{V}^{\Upsilon\eta_b}_{21}|_F$ is given in Sec.~\ref{sec:ErrVF}. The form factor inferred from experimental data in Section \ref{sec:DecayRate} is  ${V}^{\Upsilon\eta_b}_{21}|_{\text{exp}} = 0.069(14)$ and has a relative error of $19.74\%$. }
\label{tab:ErrorBudgetUps}
\begin{center}
 \begin{tabular}{ld }
    \hline \hline
    Error \% &  \multicolumn{1}{c}{~~$V^{\Upsilon\eta_b}_{21}$~~}  \\ \hline \hline 
    Systematic $\tilde{V}^{\Upsilon\eta_b}_{21}|_F$  &  10.36  \\ 
    Stats in ${V}^{\Upsilon\eta_b}_{21}$  & 5.48  \\ 
    Radiative $\alpha_s^2$ in $\omega_F $ & 0.83     \\ 
    Radiative $\alpha_s$ in $\omega_{W1} $ & 4.71     \\ 
    Radiative $\alpha_s$ in $\omega_{S} $ &  2.36    \\ 
    Radiative $\alpha_s$ in $\omega_{S1} $ & 0.51     \\ 
    Mixing down in $\omega_F^{(1)}$  & 3.92     \\ 
    Missing currents & 7.08  \\ \hline
    $a_{fm}$ scale & 1.07   \\
    Experimental masses & 0.03  \\
    Priors & 4.18  \\ \hline
    Total & 15.81  \\ \hline \hline
  \end{tabular}
\end{center}
\end{table}

By studying the error budget we see that the main sources of error are
from the systematics in $\tilde{V}^{\Upsilon\eta_b}_{21}|_F$. Here, as discussed in Sec.~\ref{sec:ErrVF}, the main sources of
uncertainty come from the statistical error in $c_4^{(1)}$ and from not
knowing the coefficient of $\alpha_s^2$ in the expansion of
$c_4$. While the statistical error on $c_4^{(1)}$ could potentially be reduced
from $7\% - 10\%$ to $2\%-3\%$ \cite{Hammant:2013}, computation of
the two-loop coefficient of $\alpha_s^2$ would be difficult and lengthy,
and unlikely to be done in the near future. Alternatively, one could
tune $c_4$ against the hyperfine splitting on all ensembles, as shown in Section \ref{sec:HyperTest}, and the error on ${V}^{\Upsilon\eta_b}_{21}$ could be reduced to $\sim 11\%$. 

After this, the main uncertainty comes from the missing currents. These
could be included with more computational time if neccessary. While the statistical error on each
current alone is around $3\%$, these statistical errors do not allow the correlations between the data
points in the fit to constrain the final result as much as we would
like, and the final error from statistics in the error budget is $5\%$
as a result. Reducing the error from statistics is unlikely to have a
sizable effect. 

Based on our analysis, we estimate that by including the next order of relativistic corrections to the current, the mixing down in $\omega_F^{(1)}$, and tuning $c_4$ against the hyperfine splitting on all ensembles, an error on $V^{\Upsilon\eta_b}_{21}$ of $8\%$ could be possible (compared against an error of $19\%$ on the value inferred from experiment), where the uncertainties in order of dominance would be from the one-loop corrections to $\omega_{W1}$ and $\omega_{S}$  and the systematic error coming from $\tilde{V}^{\Upsilon\eta_b}_{21}|_F$.

Our final answer for the form factor is:
\begin{align}
V^{\Upsilon\eta_b}_{21}(q^2 = 0) = 0.081(13) \label{eqn:FinalFormNum}
\end{align}
Final values for the decay rate and branching fraction are given in
Section \ref{sec:Conclusions}. 

\begin{figure}[t]
  \centering
  \includegraphics[width=0.45\textwidth]{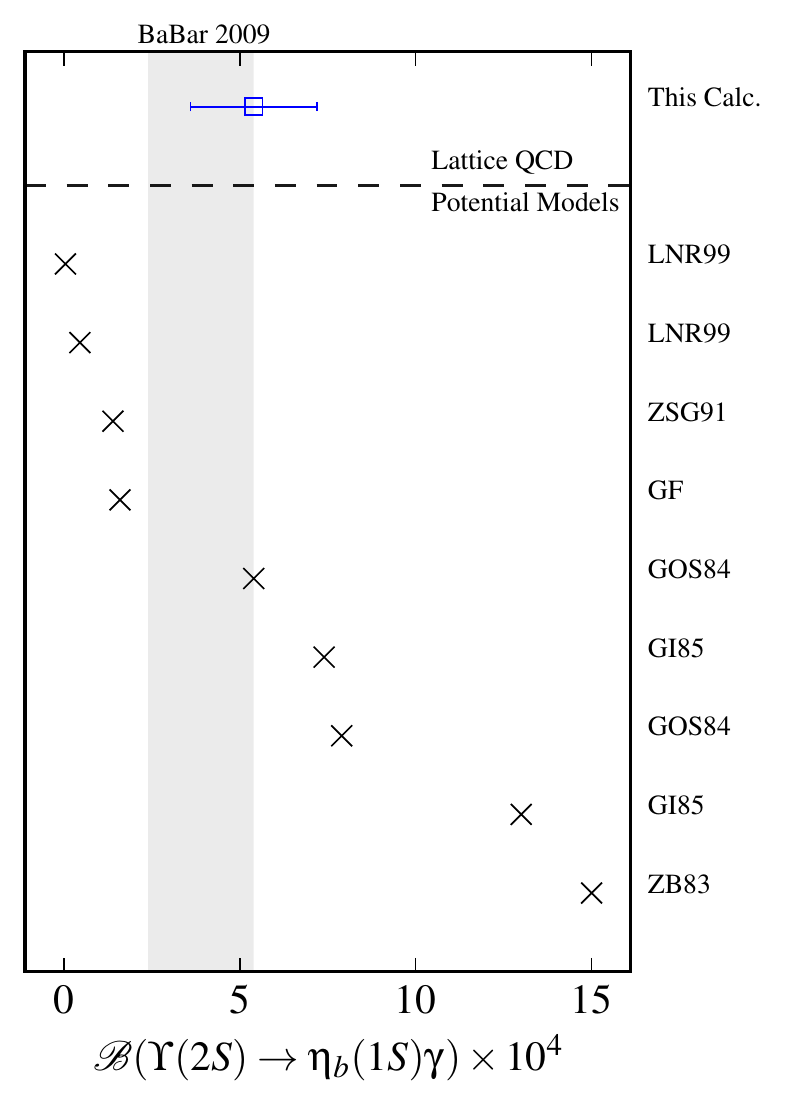}
  \caption{Comparison of our result for the branching fraction 
    (square) with experiment (vertical gray
    band) and potential model estimates from \cite{Godfrey:2001}
    (crosses). The y-axis labels the different references
    \cite{Potential:Ups2SEta1S1,Potential:Ups2SEta1S2,Potential:Ups2SEta1S3,Potential:Ups2SEta1S4,Potential:Ups2SEta1S5}
    and more information about these can be found in
    \cite{Godfrey:2001}. Using the pNRQCD decay rate
    \cite{Segovia:M1}, combined with the experimental total width from
    the PDG average given in Section \ref{sec:DecayRate}, gives a branching
    fraction of $1.9^{+8.1}_{-1.9}\times 10^{-4}$. }
  \label{fig:ComparePotLatt}
\end{figure}

\section{Discussion and Conclusions}
\label{sec:Conclusions}

In this paper we have computed the hindered M$1$
$\Upsilon(2S)\to\eta_b(1S)\gamma$ decay rate using a lattice NRQCD
formalism for the $b$-quark. We include several 
improvements on earlier exploratory work \cite{Lewis:Rad1, Lewis:Rad2} which are
fundamental to obtaining an accurate value for this decay rate. The key improvements are:
\begin{itemize}

  \item Previous work  only had one lattice spacing. We use five
    ensembles with a fully $\mathcal{O}(\alpha_s a^2)$  
    tadpole-improved L\"{u}scher-Weisz gluon action with HISQ $u,d,s$ and $c$ quarks in the sea, provided by the
    MILC collaboration. These ensembles each have $\sim 1000$
    configurations and one has physical light quark masses. 

  \item We use three relativistic corrections to the leading order
    current as described in Section \ref{sec:Currents} and we also test 
    the sensitivities of the form factors from all these currents to the
    parameters in our action as shown in Figure \ref{fig:Varyci}. 

  \item We use $\mathcal{O}(\alpha_s)$ correct values for the matching
  coefficients in the NRQCD action. We also take into account issues in tuning the $b$-quark mass
  as described in Section \ref{sec:NRQCD}. As shown in Figure
  \ref{fig:Varyci}, this decay is very sensitive to a subset of
  these parameters.

\item We calculate the $\mathcal{O}(\alpha_s)$ contribution to the
  matching coefficient of the leading order ${\psi^{\dagger}\VEC{\sigma \cdot
      B}_{\text{QED}}\psi}/2m_b$  current which mediates this decay, as described in Section \ref{sec:MatchingCoeff}.

\item While previous work extracted the matrix element by
  extrapolating/interpolating to the 
  $|\VEC{q}|_{\text{phys}}$ point, which only gives the photon on-shell
  contribution $q^2=0$ if the hyperfine splitting is correct, we use twisted boundary conditions to
  extract the form factor relevant to this decay at the physical
   $q^2=0$ point.

\end{itemize}

In Section \ref{sec:EnergyEigenstates} we performed an analysis of the  energy eigenstates of NRQCD at non-zero momentum. This is necessary as the energy eigenstates of a rotationally invariant theory, like NRQCD, in an infinite volume continuum at non-zero momentum are
  classified by helicity, unlike in a Lorentz invariant theory where
  they are described by the standard angular momentum $J$. This has important
  consequences for a lattice NRQCD calculation as additional states appear in the spectrum at non-zero
  momentum  (see Figure \ref{fig:EnergyEta}) and one has to be careful to ensure that the correct
  matrix elements are extracted from the correlator data. 

In Section \ref{sec:DecayNRQCD}, we show results for the four form factors from the currents
listed in Section \ref{sec:Currents} which when renormalised, summed and extrapolated to the
continuum limit, can be compared to the form factor inferred from experimental data. We found that
relativistic corrections to the leading order current gave a negative contribution causing destructive interference, that the power counting of the currents deviated from
what one would naively expect in NRQCD,
and that a range of sensitivities needed to be explained. 

In Section \ref{sec:PertAnalysis}, using a simple potential model, we
explained that the matrix element of the
leading order current was suppressed due to the orthogonality of the
radial wavefunctions, and this causes the matrix element to become sensitive to a multitude of effects such as
relativistic corrections to the leading order current, and certain parameters in the NRQCD action that give
rise to perturbing potentials causing relativistic corrections to
the wavefunctions, particularly those which effect the hyperfine
splitting. 

It has been suggested \cite{Lewis:Rad1, Lewis:Rad2} that the large
changes experienced in going from an unimproved calculation to an
improved calculation may mean that it would be beneficial to avoid
nonrelativistic approximations. We come to a different conclusion and illustrate that although such a calculation is intrinsically difficult, NRQCD does indeed show that a systematic approach works while also giving insight into the process under study. 

After performing the continuum and sea quark mass extrapolation, we
obtain the form factor $V^{\Upsilon\eta_b}_{21}(0)|_{\text{lat}} = 0.081(13)$, with a full error
budget in Table \ref{tab:ErrorBudgetUps}. This form factor  can be combined
with the experimental masses used in Section \ref{sec:DecayRate} to produce the decay rate:
\begin{align}
 \Gamma^{\text{lat }}(\Upsilon(2S)\to\eta_b(1S)\gamma) &= 1.72(55) \times 10^{-2} \text{ keV }  
\end{align}
which can be compared against the experimental decay rate $\Gamma^{\text{exp}}(\Upsilon(2S)\to\eta_b(1S)\gamma)$ $= 1.25(49) \times 10^{-2} \text{ keV}$ \cite{BaBar:Upsilon2S,PDG:2014}. Using the experimental total width from the PDG average given in Section \ref{sec:DecayRate} with our decay rate gives  a branching fraction of $\mathcal{B}(\Upsilon(2S)\to\eta_b(1S)\gamma) = 5.4(1.8) \times
10^{-4}$  which can be compared against the BaBar result of $3.9(1.5) \times
10^{-4}$ \cite{BaBar:Upsilon2S}. A comparison of our calculation with potential
model results including relativistic corrections \cite{Godfrey:2001}
is shown in Figure \ref{fig:ComparePotLatt}. 

Potential model predictions of  hindered M$1$ decay rates  are known to be
particularly difficult to pin down \cite{QWG:2004} and can mischaracterise
the experimental data by an order of magnitude without relativistic
corrections \cite{QWG:2010}. According to the Quarkonium Working Group
reviews \cite{QWG:2004, QWG:2010}, sources of uncertainty that contribute to
making such decays complicated to calculate include the form of the
long range potential chosen, and the  results depending  explicitly on the
quark mass and the perturbative potential chosen. 
Without relativistic corrections, the branching fraction of the
$\Upsilon(2S)\to\eta(1S)\gamma$ decay from potential model predictions ranges from $(0.67-11.0)\times 10^{-4}$ \cite{Godfrey:2001}. Due to the suppression mentioned above, the value of the decay rate is very dependent on good knowledge of the
relativistic corrections \cite{Godfrey:2001}. Including relativistic corrections,
potential model predictions for the same branching fraction  have a wider range $(0.05-15.0)\times 10^{-4}$, showing indeed that the decay rates may be sensitive to small details of
the potential \cite{Godfrey:2001}. 

The $\Upsilon(2S)\to
\eta_b(1S)\gamma$ decay is sensitive to many of the same effects as
the hyperfine splitting and an accurate calculation of this decay
relies on having the correct hyperfine splitting. Given the large
range of estimates of the hyperfine splitting from potential model
predictions ($46-87$ MeV \cite{QWG:2004}), we should not be surprised
that the potential model estimates for this decay rate also have a large range.

Additionally, radiative transitions between
bottomonium states provide a search for new-physics effects, seperate
from the weak-sector searches common in the literature
\cite{PDG:2012}. For example, the hyperfine splitting between the $\Upsilon(1S)$  and $\eta_b(1S)$ has
been an important quantity in bottomonium physics, being historically difficult for
both experimentalists and theorists to predict reliably. Using
hindered M$1$ decays, the BaBar \cite{BaBar:Upsilon3S,
  BaBar:Upsilon2S} and CLEO \cite{CLEO:Upsilon3S} experiments inferred this hyperfine splitting to be 
$\Delta^{\text{exp}.}_{\text{M}1} =  69.3 \pm
2.8$ MeV \cite{PDG:2010}.  However, in 2012, BELLE measured the
$h_b(2P,1P)\to\eta_b(1S)\gamma$ branching fractions (called E$1$
decays in the literature), removing the
dependence on hindered M$1$ decays and used a significantly larger sample of events,
 to yield a hyperfine splitting of $\Delta^{\text{exp.}}_{\text{E}1} =
57.9\pm 2.3$ MeV \cite{BELLE:hb},  where 
$\Delta^{\text{exp.}}_{\text{M}1} - \Delta^{\text{exp.}}_{\text{E}1}$
has a $3.2\sigma$ tension with being zero. 

It has been suggested
that the tension of $\Delta^{\text{exp.}}_{\text{E}1}$ and theory
\cite{Dowdall:Hyperfine} with $\Delta^{\text{exp.}}_{\text{M}1}$ could, if it persists,
indicate a hint at new physics \cite{Francesco:1S, Domingo:2009}. For
example, in a multiple-Higgs extension to the standard model, one
would speculate that the $\eta_b^{\text{exp.}}$ seen in experiments is
actually an admixture of the true $\eta_b$ and a CP-odd Higgs boson
with mass ranging from $9.4-10.5$ GeV. A relatively light CP-odd Higgs
scalar can appear in non-minimal supersymmetric extensions of the
standard model, such as the next-to-minimal supersymmetric standard
model \cite{Domingo:2009}. In such cases, the measured decay rate for
$\Upsilon(2S)\to\eta_b(1S)\gamma$ would likely differ from the
Standard Model prediction.  As stated above, this decay is sensitive
in much the same way as the hyperfine splitting. To observe a similar tension between theory and experiment here as
that existing between $\Delta^{\text{exp.}}_{\text{E}1}$ and
$\Delta^{\text{exp.}}_{\text{M}1}$ would require a $5\%$ uncertainty
on the form factor ($\sim 10\%$ on the decay rate). The error on the lattice form factor could be reduced to $\sim 8\%$ (as discussed in Section \ref{sec:fullerror})  if more precise experimental results became available. Any hint of new physics arising from a deviation between the  experimental $\Upsilon(2S)\to\eta_b(1S)\gamma$ decay rate and theory could then be explored more concretely. Additionally, the $\eta_b(2S)\to\Upsilon(1S)\gamma$ decay is an alternative approach to studying such effects and a study of this decay rate is already underway. 

E$1$ radiative decays are more easily computed than hindered M$1$ decays, 
and so the E$1$ decay rates $h_b(1P)\to\eta_b(1S)\gamma$ and $h_b(2P)\to\eta_b(1S)\gamma$ could be calculated within this NRQCD framework. Additionally, E$1$ currents can be
readily  renormalised nonperturbatively.  Combined with the experimental branching
fraction of these decays \cite{BELLE:hb}, this could give a prediction of the total width of the
$h_b(1P)$ and $h_b(2P)$.

\section*{ACKNOWLEDGMENTS}

We would like to thank Christopher Thomas for the many insightful discussions on the finite-volume lattice inflight spectrum. We are also grateful to the MILC collaboration for the use of their gauge configurations. This work was funded by STFC. The results described here were obtained using the Darwin Supercomputer of the University of Cambridge High Performance Computing Service as part of the DiRAC facility jointly funded by STFC, the Large Facilities Capital Fund of BIS and the Universities of Cambridge and Glasgow.

\appendix
\section{Classification of Particle States}
\label{app:Classification}

Theoretically, particle states living in the Hilbert space are
classified by unitary irreducible representations (irreps) of the
symmetry group of a theory.  We need to consider two symmetry groups here: the Lorentz group and the continuous rotational group in three dimensions (the symmetry group of NRQCD). The standard procedure to build infinite dimensional unitary irreps of these groups is via the method of induced representations, where one considers finite dimensional unitary irreps of the little group and then uses these to build unitary irreps of the full group.

The Poincar{\'e} group  is the symmetry group of a relativistic quantum field theory, and is given by the semi-direct product of the Lorentz group and four translations. For massive irreps of the Poincar{\'e} group, the little group is $SO(3)$\footnote{At nonzero momentum we can perform a Lorentz boost back to the rest frame, ensuring the little group is the same for zero and nonzero three-momentum} \cite{Weinberg:VOL1}. Thus in a Lorentz invariant theory, massive irreps
are defined as $| {p^2}; J, M \rangle$. Note that for quarkonia these states are eigenvectors of the charge-conjugation operator and parity is also a conserved quantum number,\footnote{At nonzero momentum, these states are not eigenvectors of the parity operator, but are eigenstates of the $\hat{\Pi}$ operator defined in the text, which conserves parity. } giving the standard $| {p^2}; J^{PC}, M \rangle$ decomposition. This description classifies experimental states seen to date \cite{PDG:2014}.

In a  continuum theory that is only rotationally invariant, the analogue of the Poincar{\'e}
group is the semi-direct product of the rotational group $SO(3)$ with the
three translations. For a rotationally invariant theory with
zero momentum, the little group is also $SO(3)$ and the states are classified as 
$| \VEC{p}^2; J, M \rangle$. Thus states in a rotationally invariant theory at
rest overlap with those in a Lorentz invariant theory at rest, where again, parity and
charge conjugation are good quantum numbers in similar situations. Given that at nonzero momentum in a rotationally invariant theory  we cannot perform a Lorentzian boost to the rest frame, the little group at nonzero momentum is now different to the zero momentum
little group. The little group is now $SO(2)$ \footnote{The construction of the irreps for a rotationally invariant theory
at nonzero momentum is similar to a massless representation in a
Lorentz invariant theory.} \cite{Weinberg:VOL1}. In this case, the unitary irreps are
classified by $| {\VEC{p}}^2; \lambda \rangle$, where $\lambda$
is an eigenvalue of the helicity operator $\hat{\lambda} =
{{\hat{p} \cdot \hat{J} }/E}$. The helicity $\lambda = \lambda_0$ will get contributions
from all $J$ with $\lambda_0 \le J$. This can have important consequences
for the extracted energy spectrum in NRQCD, c.f., Figure
\ref{fig:EnergyEta} and
\ref{fig:EnergyUps}, and is fundamentally different from
the Lorentzian theory. 

At zero momentum, the operators $i\gamma^5$ and $\gamma^i$ that we use in this calculation overlap onto $0^{-+}$ and $1^{--}$ states in a rotationally invariant continuum theory \cite{Thomas:Helicity}. We now need to find which helicity eigenstates these operators overlap with at nonzero momentum. The authors
of \cite{Thomas:Helicity} illustrate how to construct helicity operators via
\begin{align}
\mathbb{O}^{J,\lambda}({\VEC{p}}) & = \sum_{M}
\mathcal{D}^{J*}_{M\lambda}(R)\mathcal{O}^{J,M}({\VEC{p}}) \label{eqn:WignerD}
\end{align}
where $\mathcal{D}^{J}_{M\lambda}(R)$ is a Wigner-$\mathcal{D}$
matrix, $R$ is the active transformation which rotates $(0,0,|{\VEC{p}}|)$
to ${\VEC{p}}$, $\mathbb{O}^{J,\lambda}({\VEC{p}})$ is a helicity operator
with helicity $\lambda$ in an infinite volume continuum, e.g., 
\begin{align}
\langle 0 | \mathbb{O}^{J,\lambda}({\VEC{p}})
|{\VEC{p}};J', \lambda'\rangle & =
Z^{[J,J',\lambda]}\delta_{\lambda\lambda'} \label{eqn:OverlapHelicity}
\end{align}
and we refer the reader to Ref.~\cite{Thomas:Helicity} for further details. For quarkonium, the possibile values of
$\lambda = \{ 0^+, 0^- , |1|,|2|, \ldots \}$, where the $+/-$ on the
$\lambda=0$  represent the $\hat{\Pi}$ symmetry with
eigenvalue $\tilde{\eta} \equiv P(-1)^J$ \cite{Thomas:Helicity}. 
Using the fact that the Wigner-$\mathcal{D}$ matrices with $J=0$ are
$\delta_{\lambda M}$, the  $\mathcal{O}^{\gamma^5}$,
$\mathcal{O}^{\gamma^i}$ bilinear operators which we use in
this calculation give rise to the helicity operators at nonzero momentum
\begin{align}
\mathbb{O}^{J=0,\lambda=0^-}({\VEC{p}}) & =
\mathcal{O}^{\gamma^5}({\VEC{p}}) \nonumber \\
\mathbb{O}^{J=1,\lambda=0^+}({\VEC{p}}) & = \sum_M
\mathcal{D}^{J=1*}_{M\lambda=0}(R)\mathcal{O}^{\gamma^M}({\VEC{p}})
\nonumber \\
\mathbb{O}^{J=1,\lambda=|1|}({\VEC{p}}) & = \sum_M
\mathcal{D}^{J=1*}_{M\lambda=|1|}(R)\mathcal{O}^{\gamma^M}({\VEC{p}}) \,. 
\label{eqn:HelicityOps}
\end{align}
As can be seen, $\mathcal{O}^{\gamma^5}({\VEC{p}})$ is a helicity
operator which creates a $\lambda=0^-$ state, but
$\mathcal{O}^{\gamma^i}({\VEC{p}})$ creates an admixture of
$\lambda = 0^+, |1|$ states. 

The question now is: how do we identify which $J^{PC}$ contributes to
each $\lambda$, and how do we parameterise the amplitudes? By noticing
that the helicity $\hat{\lambda} = J_z$ when the momentum is projected
onto the $z$-axis, all states
with $J\ge\lambda$ will have a $J_z$ large enough to give a
contribution to this helicity state (see Table \ref{tab:Operators}). 

We also want to know how to quantify the amplitudes. In a rotationally invariant
theory, the invariant quantities are $\delta_{ij}$ and
$\varepsilon_{ijk}$. For a $J^{P}$ state, we also have the momentum
$p_J^i$ and the symmetric polarisation tensor $\epsilon^{i_1, \ldots, i_J}$. We can use these to parameterise the amplitudes relevant for a rotationally invariant theory.  For the operator $\mathcal{O}^{\gamma^i}$, Table XI in \cite{Thomas:Helicity} has the possible decompositions and we reproduce the parameterisations for the $\mathcal{O}^{\gamma^5}$ operator which are important for our calculation
\begin{align}
\langle 0 | \mathcal{O}^{\gamma^5}({\VEC{p}}) | n0^{-+}(p) \rangle
& = Z_n  \label{eqn:OverlapRotational} \\
\langle 0 | \mathcal{O}^{\gamma^{5}}({\VEC{p}}) | n1^{++}(\epsilon,p) \rangle
& = Z'_n \epsilon_i p_i /m_{n1^{++}}  \nonumber \\
\langle 0 | \mathcal{O}^{\gamma^{5}}({\VEC{p}}) | n2^{-+}(\epsilon, p) \rangle
& = Z_n^1 \epsilon_{ii} + Z^2_n\epsilon_{ij}p_ip_j /m^2_{n2^{-+}} \,.  \nonumber 
\end{align}
where $n$ is the radial label. Using the overlap for the $1^{++}$ from (\ref{eqn:OverlapRotational})
to parameterise the continuum two-point correlator with nonzero momentum, one finds that the
amplitudes from our fit with local smearing should be
suppressed by  
$|{\VEC{p}}|/m_{1^{++}}$ relative to states which overlap with the
operator at zero momentum. For the momentum that we use in our
calculation, this factor is around $7\%$, and we observe that in
our correlator data, the amplitudes for the states which do not
overlap at zero momentum (and for which we get a signal) such as the $1^{++}$, are suppressed by this factor while the other amplitudes are $\mathcal{O}(1)$. We observe that as the
momentum increases, so does the value of the amplitude at fixed lattice
spacing.

Additionally, the symmetry group giving rise to the invariants which
classify states, e.g., the little group, is broken by a finite volume lattice to a 
reduced symmetry group \cite{Moore:2005}. At zero momentum with a cubic lattice, this
reduced symmetry group for quarkonia is the octahedral group, $O_h$. States are
now classified in terms of irreps of $O_h$, denoted $\Lambda^{PC}$,
where \cite{Dudek:Excited} shows how to subduce operators with
continuum spin $J^{PC}$ to operators with definite $\Lambda^{PC}$ on
the lattice. As mentioned above, in an infinite-volume continuum theory, the
$\mathcal{O}^{\gamma^5}$ ($\mathcal{O}^{\gamma^i}$) operator
overlaps only with $J^{PC} = 0^{-+} (1^{--})$ at rest, but this
operator falls into the $A_1^{-+}$  ($T_1^{--}$) irrep of $O_h$ on the
lattice, where mixing
with the $J^{PC}=4^{-+} (3^{--}) $ state (and higher spins) is possible. However we do
not see this mixing: rotational symmetry breaking is found to be
 weakly broken  with a fine lattice and with a
rotationally invariant smearing for a particular lattice setup \cite{Dudek:Excited}, where the
spectrum and overlaps were compatible with an effective restoration of
rotational symmetry. For this reason, we choose to use a rotationally
invariant smearing, an isotropic lattice and have discretisation
improvements in our action. Secondly, the masses of the 
additional states are too large to be seen in the first few energy
levels which we are interested in. As such, they will only potentially
contribute as additional discretisation effects in the lowest energy
modes. Indeed, studies of the spectrum from NRQCD by the HPQCD
collaboration indicate this to be
the case (see Appendix C of \cite{Dowdall:Upsilon}).

For the nonzero momentum case, the reduced little group actually depends on
the type of momenta. This is due to the fact that a general integer-valued 
momentum on the lattice cannot be rotated into the
$z$-axis like in an infinite volume continuum, \footnote{With twisted
  boundary conditions, the  momenta are still discretised but just
  shifted by an arbitrary value. As such, the little group of momentum with a
  twist is the same as the little group of momentum without a twist.}
e.g.\ there is no octahedral transformation which rotates
$(0,1,1)$ to the $z$-axis. We use an isotropic momentum (rather than an
on-axis momentum) as it has been shown to
break rotational invariance less and lead to smaller discretisation
effects \cite{Dowdall:Upsilon}. For our isotropic momentum, the reduced
little group is $\text{Dic}_3$ \cite{Thomas:Helicity}. The operator
$\mathcal{O}^{\gamma^5}$ ($\mathcal{O}^{\gamma^i}$)  falls into the
$A_2$ $(A_1 \text{ and } E_2)$ irrep of $\text{Dic}_3$, where mixing
with $\lambda=3$ $(3 \text{ and } 2)$ states is possible. For
$\mathcal{O}^{\gamma^5}$, this gives rise to
potential mixing from $3^{\pm +}, 4^{\pm +}$ states (and higher spin). As in the zero-momentum case, this 
mixing due to the lattice was found to be negligible with a fine lattice and a rotationally
invariant smearing for a particular setup \cite{Thomas:Helicity}. These states should be of higher energy than the first few states in our spectrum, and we see no evidence of
them in our low lying spectrum. For the $\mathcal{O}^{\gamma^i}$ operator, there can be mixing with  $\lambda=2$ ($2\le J$ with $J_z=2$ states) which is not important for our analysis.

There is an important distinction to be understood from using a
rotationally invariant formalism for the quark versus a Lorentz-invariant
one. If each of these formalisms is discretised, then at fixed
nonzero momentum, the discretised version of the Lorentz-invariant theory might be broken to a
rotationally invariant theory, e.g., by using an anisotropic lattice spacing in
the time direction.  As such, as the infinite volume
continuum limit is taken, any overlap onto $J^{PC}$ as a result of
helicity eigenstates (such as the $1^{++}$ from the
$\mathcal{O}^{\gamma^5}$ operator) would disappear \cite{Davoudi:2012}. However, in a
rotationally invariant theory like NRQCD, as the lattice spacing is
taken to zero, these overlaps are still present as they are an infinite
volume continuum effect. This is why we find a similar signal
across all lattice spacings for these states in NRQCD.


\section{Twisted Correlators with Derivative Operators}
\label{app:TwistedCorrelators}

For clarity, we will describe the construction of the twisted
correlators with derivative operators in this section. To gain access to arbitrary momenta on the lattice, one can define a quark field \cite{TwistedBC, TwistedBC1} that satisfies $\theta$BC via $\tilde{\psi}^{\theta}(x +
e_iL)= e^{i2\pi\chi_i}\tilde{\psi}^{\theta}(x)$, where $\theta_i =
2\pi  \chi_i/L$. Now the available
momentum space is $\tilde{\Lambda} =
\{ {\VEC{k}} = {\VEC{p + \theta}} | k_i = 2 \pi (n_i + \chi_i)/L  \text{, where } n_i \in
\mathbb{Z}\}$. Notice that the available momentum space has an
arbitary shifted value $\theta$ that we can choose to give the physical
point $q^2=0$. One now builds interpolating operators from these
$\theta$BC fields as $O(x;{\theta_2\theta_1}) =
\bar{\tilde{\psi}}^{\theta_2}(x)\Gamma \tilde{\psi}^{\theta_1}(x)$,
 which gives rise to the two-point correlator
\begin{align}
&\text{C}_{\text{2pt}}({\VEC{\theta_1 - \theta_2 + p}},t)  = \sum_{{\VEC{x}}} e^{-i
  {\VEC{(\theta_1 - \theta_2 + p) \cdot x}}} \nonumber \\ 
& \hspace{1.5cm} \text{Tr}\left[ (
  \Gamma_i\tilde{S}^{\theta_2}({\VEC{0}},0| {\VEC{x}},t) ) (
  \Gamma_f\tilde{S}^{\theta_1}({\VEC{x}},t| {\VEC{0}},0) )
\right] \label{eqn:CorrelatorTwisted2pt}
\end{align}
where  $\tilde{S}^{\theta}({\VEC{0}},0| {\VEC{x}},t)$ is a quark propagator found by
inverting the Dirac matrix, $\tilde{D}^{\theta}(x,y)$, defined via $S[\tilde{\psi}^{\theta}] =
\sum_{x,y}\bar{\tilde{\psi}}^{\theta}(x)
\tilde{D}^{\theta}(x,y)\tilde{\psi}^{\theta}(y)$. As a
consequence of $\tilde{\psi}^{\theta}$ satisfying $\theta$BC, the
Dirac matrix $\tilde{D}^{\theta}(x,y)$ also satisfies the same
boundary conditions. This is an inconvenience as typical inverters are
built with PBC. However, it is possible to use a trick in
order to use the PBC invertors yet still get access to the
$\theta$BC correlator data in (\ref{eqn:CorrelatorTwisted2pt}). 

To do this, one notices that a second quark field, defined via the scaling 
$\psi^{\theta}(x) = e^{-2\pi i {\VEC{\theta \cdot x}}/L}
\tilde{\psi}^{\theta}(x)$, satisfies PBC yet still includes
information on the twist. Since
\begin{align}
\tilde{S}^{\theta}(x|y) & = e^{i {\VEC{\theta \cdot(x - y)  }}  }
S^{\theta}(x|y) \label{eqn:RelateTwistProps}
\end{align}
${S}^{\theta}(x|y)$ is a quark propagator found by
inverting the Dirac matrix, ${D}^{\theta}(x,y)$,  where $D^{\theta}(x,y)$ $=$ $e^{-i
{\VEC{\theta \cdot x  }}} \tilde{D}^{\theta}(x,y) e^{i
{\VEC{\theta \cdot y  }}}$. $D^{\theta}(x|y)$ satisfies PBC by
construction and the
two exponentials only alter the derivative in the Dirac action
and can be implemented by scaling the gluonic fields (before
inverting) as $U_{\mu}(x) \to U_{\mu}^{\theta}(x) = e^{i2\pi/L \theta_{\mu}}
U_{\mu}(x)$ with $\theta_{\mu} = (0,{\VEC{\theta}})$ \cite{TwistedBC}.

The final step is to rewrite the twisted correlator in (\ref{eqn:CorrelatorTwisted2pt}) in terms of the
propagator we actually compute using (\ref{eqn:RelateTwistProps})
\begin{align}
& \text{C}_{\text{2pt}}({\VEC{\theta_1 - \theta_2 + p}},t)  = \sum_{{\VEC{x}}} e^{-i
  {\VEC{(\theta_1 - \theta_2 + p) \cdot x}}} \nonumber \\ 
& \hspace{0.0cm} \times \text{Tr}\left[ \left(
    \Gamma_i e^{-i {\VEC{\theta_2 \cdot x}} } {S}^{\theta_2}({\VEC{0}},0| {\VEC{x}},t) \right)
    \left(( \Gamma_f e^{i {\VEC{\theta_1 \cdot x}}
    }{S}^{\theta_1}({\VEC{x}},t| {\VEC{0}},0) \right) \right] \,.
 \label{eqn:CorrelatorTwistCompute2pt}
\end{align}
If $\Gamma = \nabla$, then
\begin{widetext}
\begin{align}
\text{C}_{\text{2pt}}({\VEC{\theta_1 - \theta_2 + p}},t) & = \sum_{{\VEC{x}}} e^{-i
  {\VEC{ p\cdot x}}} \text{Tr}\big[ \left(
    e^{i {\VEC{\theta_2 \cdot x}} }\nabla_k e^{-i {\VEC{\theta_2
          \cdot x}} } {S}^{\theta_2}({\VEC{0}},0| {\VEC{x}},t) \right)
    \left( e^{-i {\VEC{\theta_1 \cdot x}} }\nabla_k e^{i
      {\VEC{\theta_1 \cdot x}} }{S}^{\theta_1}({\VEC{x}},t|
    {\VEC{0}},0) \right) \big] \,. \label{eqn:CorrelatorTwisted2ptDeriv}
\end{align}
\end{widetext}
This can be implemented in the same way as the twist in the Dirac
invertor, by using $U^{\theta}_{\mu}(x)$ in the construction of the
covariant  derivative operator. This ``changing the derivatives'' issue does not
occur in our two-point correlators, but does occur in
the (more complicated) three point correlators with currents
$J_{W1}$, $J_{S}$, $J_{S1}$ from (\ref{eqn:CurrentsKN}). To give an explicit
example of the three point correlator  using the current $J_{W1}$, by keeping the initial state at rest, and
twisting only one propagator in the
final state with $\theta_f$, we have
\begin{widetext}
\begin{align}
& {{C}}^{nm}_{\text{3pt}}({\VEC{p_f^{\theta }=p_f +
    \theta_f, q^{\theta} = q -
    \theta_f}};t,T)  = 
-i\sum_{{\VEC{x,y}}} 
e^{-i  {\VEC{ p_f \cdot x}}} 
\text{Tr}\Big[ 
  {S}^{\theta_f}({\VEC{x}},T| {\VEC{y}},t)   \nonumber \\
& \hspace{3cm} \left( \frac{e^{-i  {\VEC{ \theta_f \cdot y}}}}{8m_b^3}  \left\{
{\VEC{D}}^2, ({\VEC{ \sigma \times q^{\theta_f}}})^n e^{-i
  {\VEC{ (q - \theta_f) \cdot y}}} \right\} 
     {S}({\VEC{y}},t|{\VEC{0}},0)  \right) \sigma^m
     {S}({\VEC{0}},0| {\VEC{x}},T) 
  \Big] \label{eqn:CorrelatorTwisted3ptJW1}
\end{align}
\end{widetext}
where we can clearly see that ${\VEC{D}}^2$ does not commute with
$e^{-i  {\VEC{ \theta_f \cdot y}}}$, but not all derivatives are
twisted due to the commutation. Since there are no derivatives
in the $J_F$ current, the ${\VEC{\theta_f}}$ terms cancel and this
issue is avoided. Smearing the twisted fields leads to a
similar issue as presented above with the derivative, and so we do not smear the
twisted fields. Analogous complications arise when using point-split operators with twisted momentum in staggered quark formalisms \cite{Donald:DSPHI}. If done correctly, and any smearings are applied
appropriately, the correlator data from using
$\theta$BC and PBC should agree on a configuration basis to machine
precision (if the total momentum is identical for all states).


\section{Error Analysis Using a Simple Potential Model}
\label{app:Potential}

First, we want to find the sensitivity of the matrix element to $c_4$ using a potential from the exchange of a single gluon between two vertices involving the chromomagentic operator \cite{Hammant:2013}. We find (assuming the wavefunctions at the origin for the $\eta_b$ and $\Upsilon$ are the same)
\begin{align}
& V^{\eta_b}_{nm}  = - \frac{6c_4^2 g^2}{9m_b^2} \psi^*_n(0)
\psi_m(0)\nonumber \\
&V^{\Upsilon}_{nm}  =  \frac{2c_4^2 g^2}{9m_b^2} \psi^*_n(0) \psi_m(0) \,.
\label{eqn:c4potential}
\end{align}
Putting this back into (\ref{eqn:InteractingMatrix}) with the $J_F$ current yields:
\begin{align}
&^{(1)}\langle \eta_b(1S) | J_F | \Upsilon(2S) \rangle^{(1)}  =
\nonumber \\
& \hspace{1cm} ^{(0)}\langle \eta_b(1S) | J_F | \Upsilon(2S)
\rangle^{(0)} + \nonumber \\ 
&  \frac{c_4^2 g^2}{9m_b^2} \Bigg ( \sum_{m
  \ne 1 }\frac{6 \psi^*_1 (0) \psi_m(0)} {E_{m1}} ~^{(0)}\langle \eta_b(mS) | J_F | \Upsilon(2S) \rangle^{(0)} \nonumber \\
& - \sum_{n
  \ne 2 } \frac{ 2\psi^*_n (0) \psi_2(0) }{E_{n2}} ~^{(0)}\langle \eta_b(1S) |
J_F | \Upsilon(nS) \rangle^{(0)}  \Bigg) \,. \label{eqn:c4dependence} 
\end{align}
In getting to (\ref{eqn:c4dependence}) we have used the fact that
$E^{\Upsilon}_{nm} = E^{\eta_b}_{nm}$ as the unperturbated Hamiltonian
has no spin terms. We have neglected the
$\Upsilon(pS) \to \eta_b(1S)$ transitions for $p \ge2$ in the sum due to the fact
that the radial overlap, (\ref{eqn:WavefunctionOverlap}), is
suppressed by at least $\mathcal{O}(v^2)$. In fact, they will be suppressed more due to the
  radial difference getting larger and the wavefunction at the origin
  getting smaller for higher radial excitations. Eqn.  (\ref{eqn:c4reduced}) can be found straightforwardly by factoring the
spin part of the matrix element from the radial part, i.e., using
(\ref{eqn:WavefunctionOverlap}).

If we now consider a potential from the exchange of a single gluon involving the Darwin term at one of the vertices, we find \cite{Hammant:2013}
\begin{align}
V^{\eta_b}_{nm} = V^{\Upsilon}_{nm} =  -\frac{c_2 g^2}{3m_b^2} \psi^*_n(0)
\psi_m(0)
\,. \label{eqn:c2potential}
\end{align}
Then substituting this back into (\ref{eqn:InteractingMatrix}) we find: 
\begin{align}
&^{(1)}\langle \eta_b(1S) | J_F | \Upsilon(2S) \rangle^{(1)}  =
\nonumber \\
&\hspace{1.5cm}  ^{(0)}\langle \eta_b(1S) | J_F | \Upsilon(2S) \rangle^{(0)} \nonumber \\ 
& - \frac{c_2 g^2}{3m_b^2} \Bigg ( \sum_{m
  \ne 1 }\frac{ \psi^*_1 (0) \psi_m(0)} {E_{m1}} ~^{(0)}\langle \eta_b(mS) | J_F | \Upsilon(2S) \rangle^{(0)} \nonumber \\
& ~~~~~~+ \sum_{n
  \ne 2 } \frac{ \psi^*_n (0) \psi_2(0) }{E_{n2}} ~^{(0)}\langle \eta_b(1S) | J_F | \Upsilon(nS) \rangle^{(0)}  \Bigg ) \label{eqn:c2dependence} 
\end{align}
\begin{align}
 & = ^{(0)}\langle \eta_b(1S) | J_F | \Upsilon(2S) \rangle^{(0)} \nonumber \\ 
& - \frac{c_2 g^2}{3m_b^2 E_{21}} \psi^*_1 (0) \psi_2(0) \bigg( ~^{(0)}\langle \eta_b(2S) | J_F | \Upsilon(2S) \rangle^{(0)} \nonumber \\
& ~~~~~~~~~~ - ~^{(0)}\langle \eta_b(1S) | J_F | \Upsilon(nS) \rangle^{(0)} +
\mathcal{O}(v^2) \bigg) \,.
\label{eqn:c2reduced} 
\end{align}
Using (\ref{eqn:WavefunctionOverlap}), we see the leading order terms
in the second piece of (\ref{eqn:c2reduced}) cancel and we are left with
 $\mathcal{O}(\alpha_s v^2)$ corrections to the unperturbed matrix element. 

The four quark potential is (assuming the wavefunctions
at the origin of the two states are the same) \cite{Hammant:2013}
\begin{align}
V^{\eta_b}_{nm} =  \frac{9 d_1\alpha_s^2}{2} \frac{4}{3m_b^2}
\psi^*_n(0) \psi_m(0) \nonumber \\
V^{\Upsilon}_{nm} = \frac{9 d_2\alpha_s^2}{2} \frac{4}{3m_b^2} \psi^*_n(0) \psi_m(0)
\,. \label{eqn:4qpotential}
\end{align}
Putting this into (\ref{eqn:InteractingMatrix}) and performing an
identical analysis as done above gives
\begin{align}
& ^{(1)}\langle \eta_b(1S) | J_F | \Upsilon(2S) \rangle^{(1)}  =
\nonumber \\
& \hspace{1cm} ^{(0)}\langle \eta_b(1S) | J_F | \Upsilon(2S) \rangle^{(0)} \nonumber \\ 
& - \frac{9d_1 \alpha_s^2}{2} \frac{4}{3m_b^2} \sum_{m
  \ne 1 }\frac{ \psi^*_1 (0) \psi_m(0)} {E_{m1}} ~^{(0)}\langle \eta_b(mS) | J_F | \Upsilon(2S) \rangle^{(0)} \nonumber \\
& - \frac{9d_2\alpha_s^2}{2} \frac{4}{3m_b^2} \sum_{n
  \ne 2 } \frac{ \psi^*_n (0) \psi_2(0) }{E_{n2}} ~^{(0)}\langle
\eta_b(1S) | J_F | \Upsilon(nS) \rangle^{(0)} \nonumber \\ 
& = ~ ^{(0)}\langle \eta_b(1S) | J_F | \Upsilon(2S) \rangle^{(0)} \nonumber \\ 
& \hspace{1.0cm} + \frac{9}{2} \frac{4}{3m_b^2} \frac{ \psi^*_1 (0) \psi_2(0)}
{E_{21}}\mathcal{S}_{if} \Big(  d_2\alpha_s^2 - d_1\alpha_s^2
  \nonumber \\ 
&  \hspace{3.5cm}+ \mathcal{O}\big((2d_2\alpha_s^2 -d_1 \alpha_s^2)v^2 \big) \Big)\,. \label{eqn:4qreduced}  
\end{align}
The error in the last line was introduced
by expanding out the radial overlap (\ref{eqn:WavefunctionOverlap}) and
noting that the two matrix elements do not have to be identical to first
order in $|q_{\gamma}|^2$. Even if we did include the four fermion
operators in the calculation, since only the combination $d_1 -d_2$ is
currently known perturbatively, and not $d_1$ and $d_2$ individually, we would
still need to introduce the $\mathcal{O}(v^2)$ error in our
calculation.

\bibliographystyle{h-physrev5}
\bibliography{./CH}

\end{document}